\documentclass[10pt,journal]{IEEEtran}
\usepackage{generic}
\usepackage{cite}
\usepackage{mathtools}
\usepackage{amssymb}
\usepackage{amsmath,amssymb,amsfonts}
\usepackage{textcomp}
\usepackage{url}

 \usepackage{pgfplots}
 \pgfplotsset{compat=1.17}
 \usepgfplotslibrary{statistics}
\usepackage{listings}
\usepackage{listingsutf8}
\usepackage{xcolor}

\usepackage{listings}
\usepackage{xcolor}

\lstdefinelanguage{json}{
  basicstyle=\ttfamily\small,
  numbers=left,
  numberstyle=\tiny\color{gray},
  stepnumber=1,
  numbersep=5pt,
  showstringspaces=false,
  breaklines=true,
  frame=single,
  backgroundcolor=\color{white},
  literate=
   *{0}{{{\color{blue}0}}}{1}
    {1}{{{\color{blue}1}}}{1}
    {2}{{{\color{blue}2}}}{1}
    {3}{{{\color{blue}3}}}{1}
    {4}{{{\color{blue}4}}}{1}
    {5}{{{\color{blue}5}}}{1}
    {6}{{{\color{blue}6}}}{1}
    {7}{{{\color{blue}7}}}{1}
    {8}{{{\color{blue}8}}}{1}
    {9}{{{\color{blue}9}}}{1}
    {:}{{{\color{black}:}}}{1}
    {,}{{{\color{black},}}}{1}
    {"}{{{\color{red}"}}}{1}
    {[}{{{\color{teal}[}}}{1}
    {]}{{{\color{teal}]}}}{1}
    {{\{}}{{{\color{teal}\{}}}{1}
    {{\}}}{{{\color{teal}\}}}}{1},
}

\usepgfplotslibrary{statistics}

\usepackage{array}
\usepackage{caption}
\usepackage{subcaption}

\usepackage{graphicx}
\usepackage{multirow}
\usepackage{booktabs}

\usepackage{MnSymbol}
\usepackage{threeparttable}
\usepackage{enumerate}

\usepackage[linesnumbered, ruled, vlined]{algorithm2e}

\SetCommentSty{mycommfont}
\usepackage{caption}
\usepackage{multirow}
\usepackage{array}
\usepackage{graphicx}

\usepackage{siunitx}
\usepackage{booktabs}

\usepackage{pifont}% http://ctan.org/pkg/pifont
\newcommand{\cmark}{\ding{51}}%
\newcommand{\xmark}{\ding{55}}%

\begin{document}
\title{TINC: Trusted Intelligent NetChain}
\author{Qi Xia, 
Hu Xia, 
Isaac Amankona Obiri, 
Adjei-Arthur Bonsu, 
Grace Mupoyi Ntuala, 
Ansu Badjie,
Tienin Bole Wilfried,
Jiaqin Liu,  
Lan Ma, 
Jianbin Gao\textsuperscript{*},  
and Feng Yao    % <-this % stops a space
\thanks{

The authors are with the School of Computer Science and Engineering (School of Cyber Security), University of Electronic Science and Technology of China (UESTC), Chengdu, 611731, China, and also with College of Systems Engineering, National University of Defense Technology Changsha, 410073, China. (e-mail: xiaqi@uestc.edu.cn; xiahu@uestc.edu.cn; obiriisaac@gmail.com; peeman34@gmail.com; 202114080110@std.uestc.edu.cn; 202214080108@std.uestc.edu.cn; tieninwilfried@std.uestc.edu.cn; jqliu@uestc.edu.cn; malan@uestc.edu.cn; gaojb@uestc.edu.cn; fengyao@nudt.edu.cn).
}

}

\maketitle

\begin{abstract}
Blockchain technology facilitates the development of decentralized systems that ensure trust and transparency without the need for expensive centralized intermediaries. However, existing blockchain architectures—particularly consortium blockchains—face critical challenges related to scalability and efficiency. State sharding has emerged as a promising approach to enhance blockchain scalability and performance. However, current shard-based solutions often struggle to guarantee fair participation and a balanced workload distribution among consortium members. To address these limitations, we propose Trusted Intelligent NetChain (TINC), a multi-plane sharding architecture specifically designed for consortium blockchains. TINC incorporates intelligent mechanisms for adaptive node assignment and dynamic workload balancing, enabling the system to respond effectively to changing network conditions while maintaining equitable shard utilization. By decoupling the control and data planes, TINC allows control nodes to focus on consensus operations, while data nodes handle large-scale storage, thus improving overall resource efficiency. Extensive experimental evaluation and formal analysis demonstrate that TINC significantly outperforms existing shard-based blockchain frameworks. It achieves higher throughput, lower latency, balanced node and transaction distributions, and reduced transaction failure rates. Furthermore, TINC maintains essential blockchain security guarantees, exhibiting resilience against Byzantine faults and dynamic network environments. The integration of Dynamic Decentralized Identifiers (DDIDs) further strengthens trust and security management within the consortium network.
\end{abstract}


\section{Introduction}

Blockchain technology has emerged as a key innovation, providing a decentralized, immutable, and transparent framework for secure and verifiable transactions without reliance on trusted centralized authorities \cite{nakamoto2008bitcoin,al2018blockchain}. Originally designed to support cryptocurrencies, blockchain has since evolved into a versatile infrastructure with applications in various domains, including finance, supply chain management, healthcare, and government systems. For consortium networks, where participants are pre-approved and authenticated, blockchain provides a mechanism for multiple organizations to collaborate on a shared, trustworthy ledger, thus providing transparency and operational efficiency \cite{yao2021survey}.

Existing  monolithic blockchain architectures\cite{nakamoto2008bitcoin,wood2014ethereum} face significant scalability and performance challenges, especially in consortium environments. These traditional designs require every node to store a complete blockchain replica and participate in consensus, leading to inefficiencies like high storage needs and limited throughput \cite{do2023sok,obiri2024hiba}. Sharded architectures aim to address these issues by partitioning the blockchain into shards for parallel processing, but they introduce complexities such as cross-shard communication and synchronization delays\cite{cai2022benzene,nguyen2019optchain,wang2019monoxide,hong2022scaling,wang2019sok}.

Emerging research addresses these limitations through architectural and protocol refinements. An overlapping sharded blockchain architecture was introduced in \cite{hong2022scaling,liu2021overlapshard,xu2024overlapping}, which transforms cross-shard transactions into overlap-shard transactions. This approach enables the validation of cross-shard transactions to occur within the overlapping segments of the shards, effectively minimizing both processing and synchronization overhead. Hierarchical blockchain architectures extend the sharding paradigm by organizing shards into multiple tiers, with higher-level shards coordinating operations among lower-level shards to optimize inter-shard communication and resource distribution \cite{obiri2024hiba,kantesariya2021optishard,yang2019blockchain}. Machine learning-driven solutions, such as LB-Chain, predict transaction loads and dynamically reallocate accounts across shards to achieve load balancing \cite{li2023lb}. Similarly, community detection techniques have been applied to optimize account partitioning, balancing shard loads and minimizing cross-shard transaction overhead \cite{zhang2023txallo}. 

Additionally, transaction allocation strategies have been proposed to co-locate related transactions within the same shard, thereby minimizing inter-shard dependencies and improving confirmation latency \cite{tao2023sharding}. Elastic resource management frameworks, such as the Lyapunov-based approach in \cite{blockchains2022elastic}, dynamically allocate computational resources to shards of the consortium blockchain, ensuring stable transaction throughput even under bursty transaction injections or imbalanced load conditions.

Although existing approaches address specific challenges such as transaction throughput, load balancing, and cross-shard communication, they focus little on consortium blockchain and do not provide a comprehensive framework suitable for consortium blockchain systems. Key components, such as the decentralized public key infrastructure (DPKI) \cite{fromknecht2014decentralized}, responsible for managing participant identities and ensuring trust within the network, remain underexplored. Similarly, equitable and adaptive distribution of nodes and dynamic workload  (unprocessed transactions) allocation are inadequately addressed. For instance, traditional sharding mechanisms often lack a Decentralized Identity Management (DIM) system, essential for providing effective Dynamic Digital Identity (DDID). The absence of such a system hinders the ability to manage digital identities efficiently, making the system vulnerable to identity fraud and unauthorized access. Furthermore, trust management in permissioned blockchain networks requires an equitable allocation of nodes across shards to ensure that each consortium retains proportional representation. This balanced distribution prevents any single consortium from being sidelined in shard-level decisions and fosters inclusivity in the consensus process.

To address these challenges, we propose the trusted intelligent netchain (TINC), a  permissioned blockchain architecture tailored for consortium networks. TINC employs a hierarchical, multi-plane design to partition responsibilities across three specialized layers: the Root Plane, Control Plane, and Data Plane, ensuring modularity, scalability, and secure operation.

By decoupling responsibilities across various planes, TINC effectively mitigates the inherent limitations of current architectures, presenting a unified framework that improves scalability, trust management, and resource efficiency. In contrast to conventional static Decentralized Identifiers (DIDs) that remain immutable post-creation, TINC introduces Dynamic Decentralized Identifiers (DDIDs) that can adapt and evolve in response to changes in user features, permissions, or reputation metrics. This empowers users with complete sovereignty over their identities, eliminating the need for centralized authorities.

In summary, our contributions are as follows:
\begin{itemize}
    \item We propose a novel multi-plane architecture that divides the blockchain into three specialized planes: the Root Plane, Control Plane, and Data Plane. This modular structure isolates responsibilities to enhance scalability, improve resource management, and ensure secure operation.

    \item We design a dynamic decentralized identifier(s) architecture that provides an adaptive and real-time update to identity management on blockchain networks. 
    
    \item We design a protocol for adaptive node and workload distribution across the Control Plane and Data Plane shards. This protocol dynamically balances workloads to reduce cross-shard dependencies, mitigate bottlenecks, and optimize transaction throughput under varying conditions. Additionally, it ensures equitable distribution of nodes within each shard to prevent any shard from being overtaken by a consortium.
    
    \item We provide an implementation of the proposed scheme, along with a comparison to related works. The results demonstrate that our scheme outperforms existing solutions and offers additional functionalities.
\end{itemize}

The remainder of this paper is organized as follows. Section \ref{relatedworks} reviews related work on blockchain sharding and decentralized identifiers. Section \ref{background} provides background on blockchain technology, consensus algorithms, and sharding. Section \ref{systemDesign} details the TINC architecture, including its planes, components, and protocols. Section \ref{processFlow} presents the process flow for network setup and operation. Section \ref{securityAnalysis} provides a formal security analysis of the TINC framework. Section \ref{evaluation} describes our experimental setup and implementation. Section \ref{conclusion} concludes the paper.

\section{Related works}\label{relatedworks}
\subsection{Sharded blockchain}
Various sharding techniques\cite{blockchains2022elastic,luu2016secure,zamani2018rapidchain,wang2019monoxide,gao2024neuchain+} have emerged to address monolithic blockchain scalability. Benzene\cite{cai2022benzene} introduced cooperation-based random sharding with secure randomized assignments. OptChain\cite{Nguyen2020} improved throughput by grouping related transactions, maintaining parallelism through a balanced shard distribution. However, increased cross-shard interactions introduce significant overhead.

To reduce this overhead, an overlapped sharding was proposed. Pyramid\cite{hong2022scaling} introduced an overlap consensus, which improved intra-shard cooperation. OverlapShard\cite{liu2021overlapshard} assigned nodes to multiple shards, reducing the frequency of cross-shard. OverShard\cite{yu2023overshard} integrated virtual accounts and overlapping networks to ease inter-shard transaction bottlenecks. Despite improved scalability, these approaches underperform in resource-constrained scenarios like IIoT due to limited cross-shard optimization.

To address IIoT constraints, OSOM\cite{xu2024overlapping} optimized node assignment using properties and transaction types, enhancing throughput and reducing storage. However, it struggles with scalability in dynamic environments where inter-shard interactions remain costly. Hiba\cite{obiri2024hiba} introduced a Hierarchical High-Performance Blockchain Architecture with a novel fitness function for transaction-shard mapping, minimizing cross-shard traffic and balancing workloads. Although efficient, it lacks dynamic management of identity, resources, and real-time workload adaptation, critical for large-scale, dynamic consortium systems.

Consortium blockchain sharding remains underexplored. Meepo\cite{zheng2021meepo} uses partial cross-call merging to support smart contract dependencies, boosting throughput. Zhou et al.\cite{zhou2020dynamic} presented adaptive shard reconfiguration based on traffic and conditions. Wang et al.\cite{wang2023two} added privacy with a two-layer model combining sharding and confidential transaction support. However, access control aligned with dynamic decentralized digital identity (DDID) systems remains unaddressed.

\subsection{Decentralized Identifiers }
Decentralized Identifiers (DIDs) and Self-Sovereign Identity (SSI) are emerging paradigms that give users control over their digital identities, leveraging blockchain technology for enhanced security and privacy \cite{CedilloOrellana2023}. DIDs and Verifiable Credentials (VCs) have been standardized by W3C 
\cite{DIDCore2020, VCDataModel2019} and are applicable beyond individuals, extending to cloud, edge, power grid and IoT resources \cite{Mazzocca2024}. These technologies are being explored in various domains, including metaverse-enhanced education, which addresses privacy, security, and interoperability concerns \cite{Polychronaki2024}. While DIDs offer transparency and user control over disclosed information, the public accessibility of DID documents raises potential privacy risks, particularly through service properties that may leak sensitive information during authentication processes \cite{Kim2021}. Despite their potential, challenges remain in implementing DIDs and SSI solutions, necessitating further research to address standardization, scalability, and widespread adoption across different sectors \cite{CedilloOrellana2023} and \cite{Polychronaki2024}.

\section{Background}\label{background}

PBFT ensures both safety and liveness, meaning that all honest nodes agree on the same value while achieving consensus in a limited number of rounds, provided the leader is functioning correctly. It is particularly well-suited for consortium blockchains because it offers fault tolerance, computational efficiency, and deterministic finality. Unlike Proof of Work (PoW) and Proof of Stake (PoS), which are resource-intensive, or RAFT, which does not provide Byzantine fault tolerance, PBFT effectively combines efficiency, resilience to faults, and finality, making it an ideal choice for consortium blockchains.

\subsection{Sharding for Scalability in Blockchain}

Sharding is a technique adapted from database systems to enhance blockchain scalability by dividing the network into smaller groups called "shards." Each shard processes a subset of transactions and maintains a partial copy of the blockchain’s state. This parallel operation reduces individual node workload, increases transaction throughput, and improves scalability \cite{wang2019monoxide}. In state sharded blockchains, intra-shard transactions are processed quickly within a shard, while cross-shard transactions, which involve multiple shards, are more complex and resource-intensive due to required coordination and consensus checks. Sharding also raises security concerns, particularly in cross-shard interactions, necessitating effective node allocation and security mechanisms to protect network integrity.
\subsection{Node Allocation Protocol}
This protocol aims to assign nodes to shards in a manner that balances load, optimizes computational resources, ensures geographic diversity, and prevents potential malicious dominance. The algorithm leverages controlled randomization, periodic reallocation, and comprehensive checks on node features to create a resilient, fault-tolerant network structure.
\begin{figure}[htb!]
    \centering
    \includegraphics[width=1.0\linewidth]{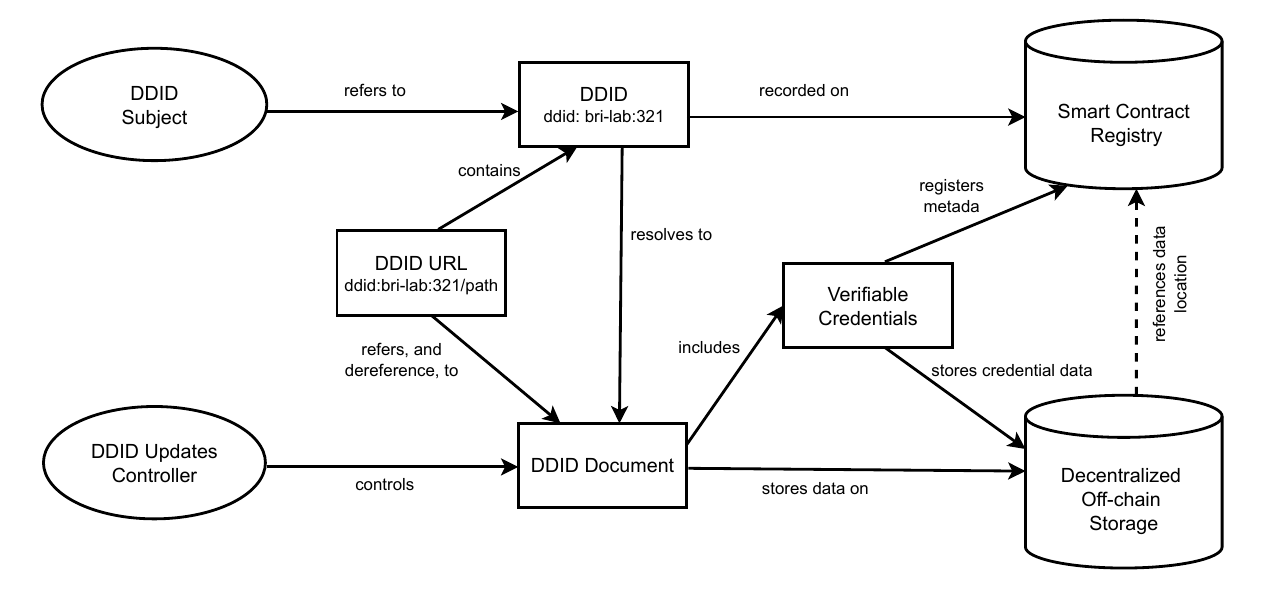}
    \caption{Overview of DDID architecture}
    \label{fig:ddid}
\end{figure}
\subsection{Dynamic Decentralized Identifier}
Decentralized Identifiers (DIDs) serve as a foundation for self-sovereign identity management in distributed systems, allowing for verifiable digital identities without relying on centralized registries or identity providers \cite{did_core,vc_data_model}. According to W3C standards, a DID signifies any subject defined by its controller \cite{DIDCore2020}. 

While DIDs enhance trust in blockchain settings, traditional implementations tend to be static, making them unsuitable for consortium blockchains where trust and roles are dynamic \cite{Kim2021}. To address this, we propose Dynamic Decentralized Identifiers (DDIDs), which introduce controlled mutability to identity management. DDIDs allow specific attributes to evolve while maintaining cryptographic verifiability and an immutable audit trail, essential for networks with changing organizational roles \cite{yin2025dp}.

The DDID protocol in TINC employs a threshold signature scheme based on BLS \cite{bacho2022adaptive} for efficient multi-party authorization of identity updates. This approach ensures that no single entity can unilaterally modify identity attributes, aligning with consortium governance principles. The threshold signature mechanism requires that at least $t$ out of $n$ authorized controllers must approve any DDID update, providing both security and operational flexibility. As shown in Fig.\ref{fig:ddid}, the DDID architecture consists of the following components:

\begin{itemize}
    \item DDID Subject: The entity (node, organization, or service) represented by the identifier within the consortium blockchain. This typically corresponds to a node operating in one of TINC's three planes.
    
    \item DDID Document: A machine-readable document containing all information associated with the DDID, structured as a JSON-LD document following the W3C DID specification with custom extensions for temporal attributes. The document stores:(a) \textit{Authentication Keys}: Ed25519 public keys \cite{brendel2021provable} for signature verification; (b) \textit{Update Control Rules}: Policies specifying which entities can approve changes; (c) \textit{Temporal Attributes}: Time-bound permissions with validity periods; (d) \textit{Version History}: Merkle-based proof tree \cite{merkle1980protocols} of all document versions.
    
    \item Dynamic Updates Controller (DUC): A smart contract enforces update rules using the threshold signature verification algorithm. Update requests follow a two-phase commit protocol: (a) \textit{Proposal Phase}: The requested change is broadcast to all authorized controllers; (b) \textit{Approval Phase}: At least $t$ controllers sign the proposal with their private keys; (c) \textit{Execution Phase}: When threshold is reached, the DUC verifies the aggregate signature and commits the change.
    
    \item Smart Contract Registry: A blockchain-based registry that maintains the current state of all DDIDs using a combination of Patricia trees \cite{szpankowski1990patricia} for efficient lookup and zero-knowledge proofs for selective disclosure of attributes \cite{flamini2024cryptographic}.
    
    \item Decentralized Off-chain Storage: A content-addressable storage system for larger DDID-related data using a Distributed Hash Table (DHT) with on-chain hash anchors, inspired by IPFS \cite{chen2022scalable}.
    
    \item Verifiable Credentials: Cryptographically secure, portable claims about the DDID subject following the W3C Verifiable Credentials Data Model \cite{VCDataModel2019}, with NIZKP-based selective disclosure mechanisms \cite{flamini2024cryptographic}.
\end{itemize}

\subsubsection{DDID Implementation in TINC's Sharding Architecture}

Within TINC's sharded architecture, DDIDs serve three essential functions that directly support the system's core objectives:

\begin{enumerate}
    \item Shard-Specific Authorization: DDIDs encode shard-specific permissions that determine which validation, consensus, and storage operations a node can perform within each shard. These permissions are dynamically updated as nodes are redistributed across shards during rebalancing operations, ensuring seamless adaptation to changing network conditions.
    
    \item Equitable Consortium Representation: Each node's DDID includes a consortium affiliation attribute that is used during the node distribution process to enforce the equitable representation constraint defined in Equation \eqref{eq:consortium_balance}. This prevents any single consortium from dominating any particular shard.
    
    \item Dynamic Trust Scoring: Nodes accumulate reputation scores based on their historical behavior (transaction validation accuracy, uptime, and consensus participation). These scores, cryptographically protected in the DDID document using Verifiable Credentials \cite{VCDataModel2019}, influence future node allocation decisions and permission assignments.
\end{enumerate}

Fig. \ref{fig:ddid} provides a sample DDID implementation in our system for \emph{ddid:bri-lab:f2e8d7c9b3a4}, where "bri-lab" identifies the organization and "f2e8d7c9b3a4" is a unique cryptographic identifier.

\begin{figure}[!ht]
    \centering
    \includegraphics[width=6cm]{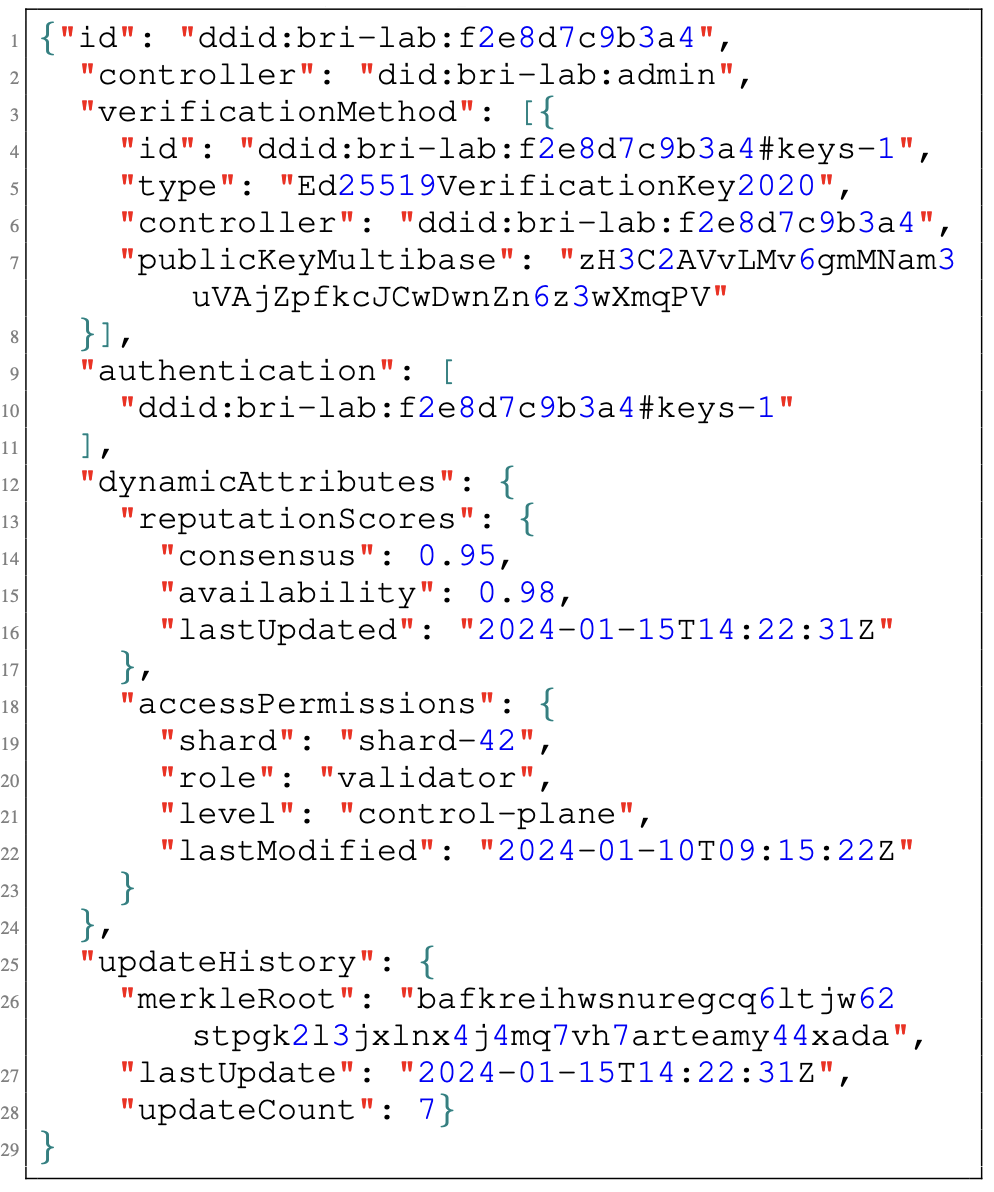}
    \caption{Sample DDID Document}
    \label{fig:did}
\end{figure}
This DDID document demonstrates how temporal attributes like reputation scores and access permissions are maintained with timestamps to track their evolution, enabling TINC to make informed decisions about node allocation and permission assignment. Table \ref{Tab:DDID} illustrates the key advantages of DDIDs over static DIDs \cite{did_core,vc_data_model} specifically in the context of sharded consortium blockchains, where the ability to adapt to changing node roles and consortium membership is essential.

\begin{table}[ht]\scriptsize
    \centering
    \caption{Comparison: Static DID vs. Dynamic DID (DDID) for Consortium Blockchains}\label{Tab:DDID}
    \begin{tabular}{|m{2.5cm}|m{2.2cm}|m{2.8cm}|}
    \hline
        {Feature} & {Static DID \cite{did_core,vc_data_model}} & {Dynamic DID (DDID)} \\
        \hline
        Node Reallocation & Requires new DID creation & Seamlessly updates permissions \\
        \hline
        Consortium Representation & Fixed attribution & Adaptive representation tracking \\
        \hline
        Trust Management & Binary trust model & Continuous reputation scoring \\
        \hline
        Permission Granularity & Coarse-grained & Shard-specific permissions \\
        \hline
        Temporal Constraints & Not supported & Time-bound authorizations \\
        \hline
    \end{tabular}
\end{table}

By integrating DDIDs into our sharded architecture, TINC achieves a key innovation in consortium blockchain management: the ability to dynamically adjust node permissions and shard assignments without disrupting network operations. This capability directly supports our core objectives of equitable consortium representation, adaptive workload distribution, and secure cross-shard communication, making DDIDs an essential component of our overall architecture rather than a separate feature.
\section{System Design}\label{systemDesign}

\begin{figure}
    \centering
    \includegraphics[width=7.5cm]{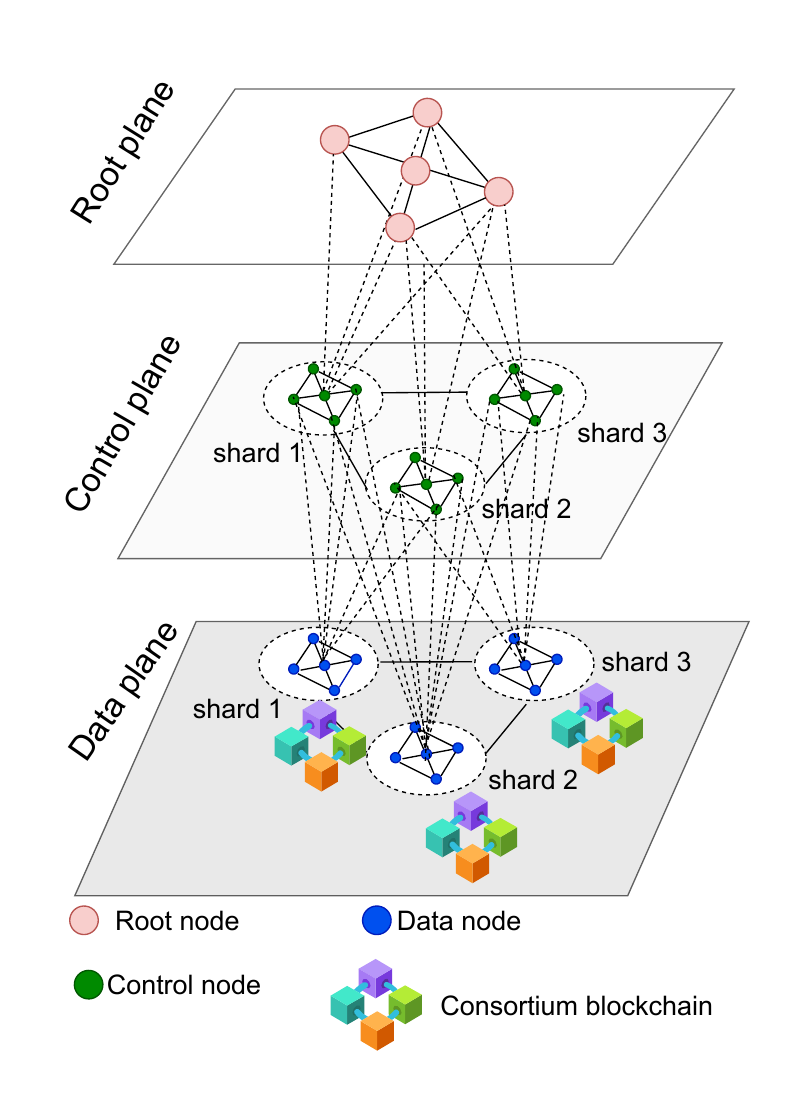}
    \caption{System Architecture}
    \label{fig:layerview}
\end{figure}

\begin{figure}
    \centering
    \includegraphics[width=8.5cm]{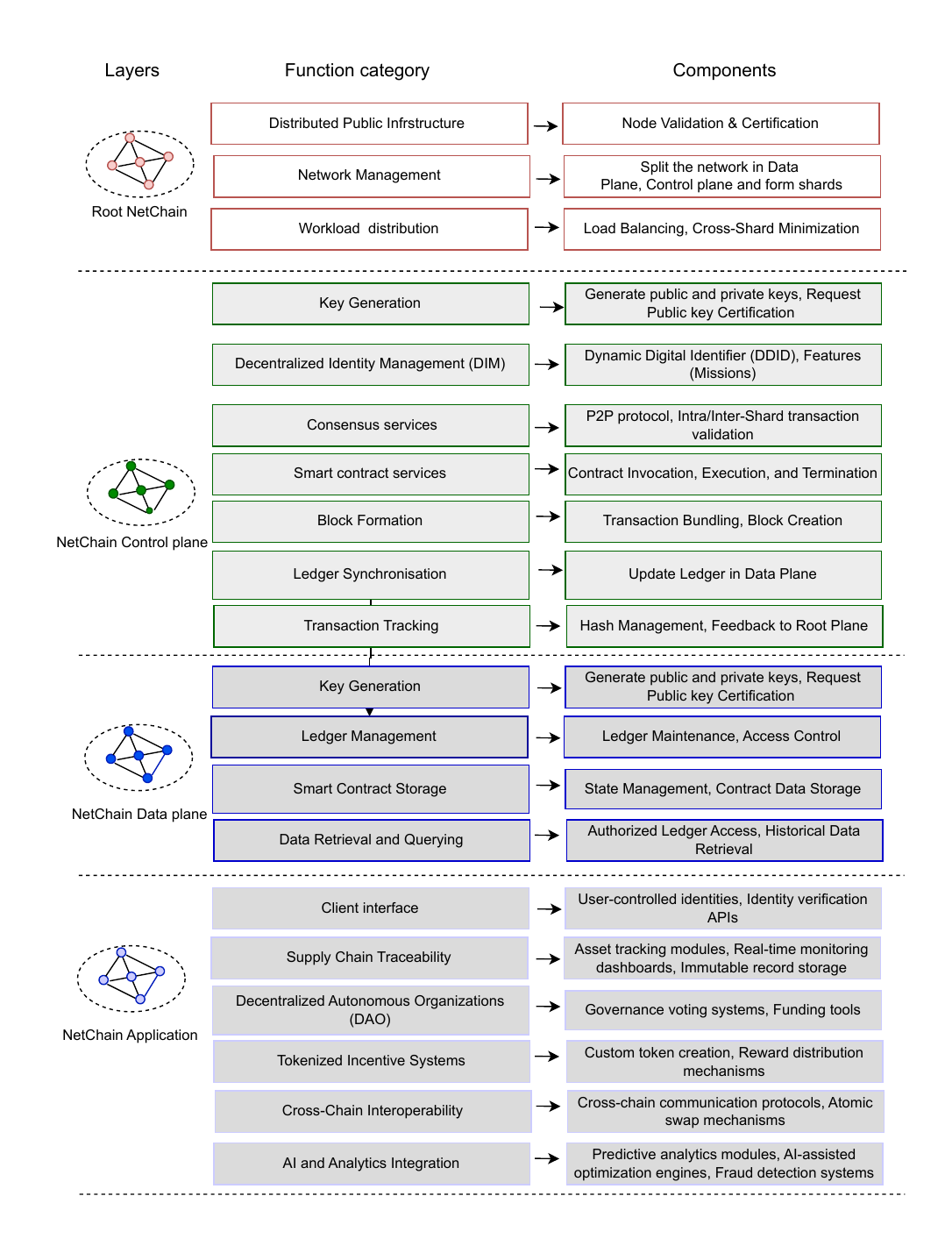}
    \caption{Layer view of functionality}
    \label{fig:archtecture}
\end{figure}

As shown in Figures \ref{fig:layerview} and \ref{fig:archtecture}, the proposed model consists of three planes, namely root plane, control plane, and data plane. The functionality of each of the plane is explained as follows: 
\begin{itemize}
     \item Root plane: The Root Plane handles key security and management tasks for the network using a Distributed Key Generation (DKG) scheme for Private Keys (SK\textsubscript{i}) and threshold keys (TSK\textsubscript{i} and TPK\textsubscript{i}). It ensures node validation through signed certificates with TSK\textsubscript{i}, allowing only authenticated nodes. It optimizes node allocation across shards for load balancing and minimizes cross-shard dependencies by grouping related transactions in the same shard. The Root Plane also oversees Smart Contract Deployment and access control within the network.
      \item Control Plane: The Control Plane oversees transaction and consensus processes within and between shards, ensuring network consistency through validation mechanisms for contract-based transactions. It manages smart contract life cycles, including invocation and execution. Block Formation consolidates validated transactions into finalized blocks, while Ledger Synchronization updates the Data Plane ledger. Transaction Process Tracking includes managing transaction hashes and states and provides feedback to the Root Plane for new transaction sets.
     \item Data Plane: The Data Plane is responsible for maintaining the ledger and handling smart contract data. Ledger Management ensures an up-to-date record of all validated transactions, and Smart Contract Storage provides persistent storage for contract states and data. Data Retrieval \& Querying manages authorized access to the ledger, allowing retrieval of historical data while ensuring data privacy and security. The Data Plane thus enables efficient data management and secure access to transaction records across the shard system.
\end{itemize}

\section{Process Flow}\label{processFlow}
\begin{figure*}
    \centering
    \begin{minipage}{0.5\textwidth}
        \centering
        \includegraphics[width=10cm]{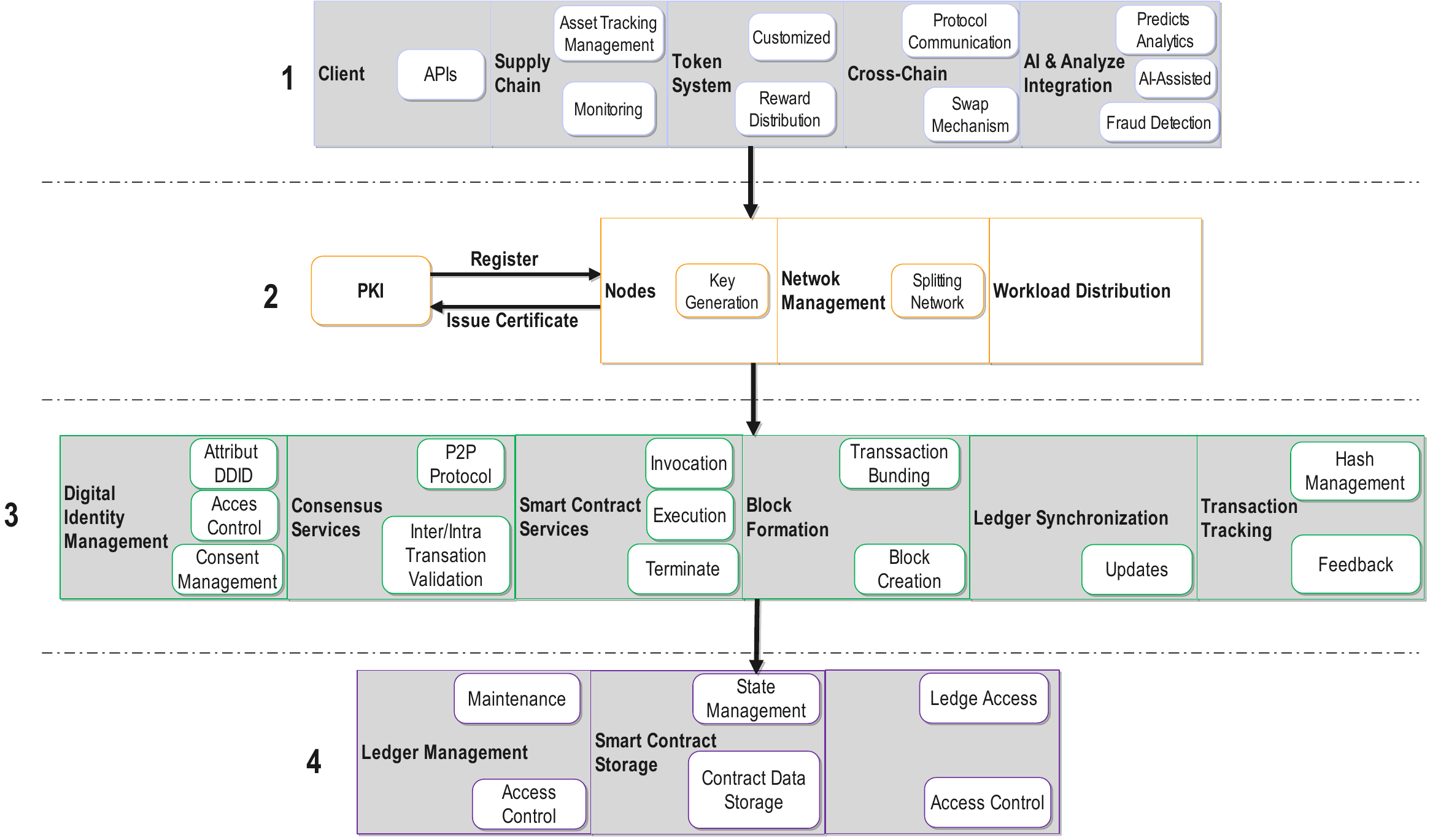}
        \caption{TINC Framework}
        \label{fig:TINC Framework}
    \end{minipage}%
    \begin{minipage}{0.5\textwidth}
        \centering
        \includegraphics[width=7cm]{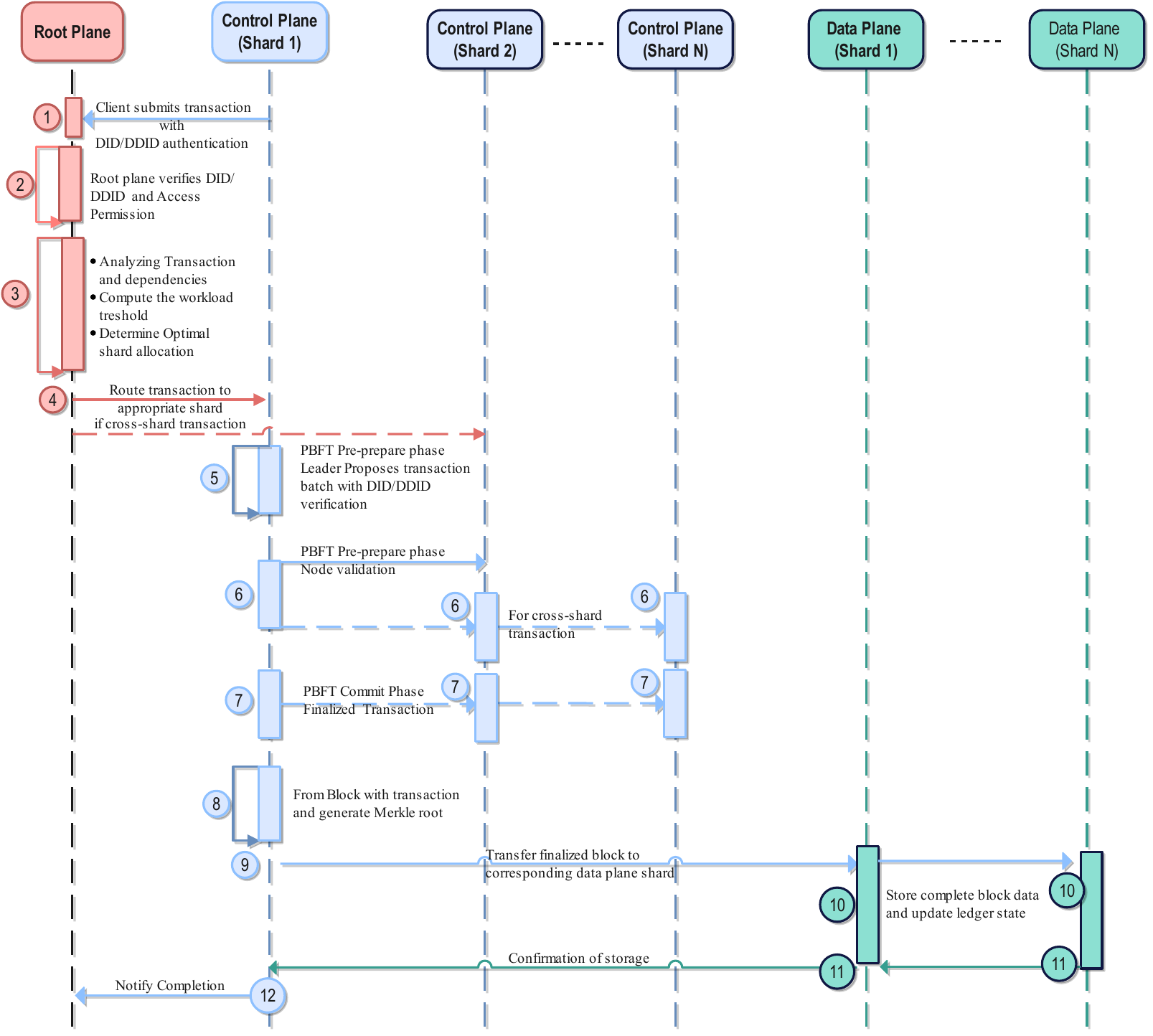}
        \caption{TINC Flow Diagram}
        \label{fig:TINC Flow Diagram}
    \end{minipage}
\end{figure*}

The proposed scheme is divided into six key phases: Network Setup, DDID Integration within Control Plane, Node Distribution, Workload Distribution, Transaction Processing and Commitment with DDID Verification, and DDID Management and Revocation, as shown in Fig.\ref{fig:TINC Framework}. These phases are designed to ensure a robust, scalable, and secure blockchain network architecture that supports high throughput and optimal performance while maintaining equitable consortium representation through dynamic identity management. The detailed flow diagram is illustrated in Fig.\ref{fig:TINC Flow Diagram}.

\subsection{Network Setup}

The network setup phase begins with the root nodes establishing a Decentralized Public Key Infrastructure (DPKI) to verify the identities and public keys of the control and data nodes. Root nodes represent the authoritative entities of the consortium blockchain, tasked with bootstrapping the network and ensuring its foundational security. Within the system, there are two sets of IDs: static DIDs and dynamic DIDs (DDIDs). DDIDs of an entity can progressively be obtained during interaction with the system. Users within the consortium, equipped with identity credentials issued by their respective organizations, generate a public-private key pair. They then submit their public key to the root nodes for certification. The root nodes validate the user's identity and verify their possession of the corresponding private key using NIZK \cite{hao2017schnorr}. Upon successful verification, the root nodes issue a certificate to the user based on a decentralized PKI model. This certificate serves as the user's proof of authenticity within the network.

\subsection{DDID Integration within Control Plane}

After the initial node authentication and allocation to appropriate planes by the root plane, nodes in the control plane transition to using Dynamic Decentralized Identifiers (DDIDs). The DDID system is managed by the Decentralized Identity Manager (DIM) component within the control plane, enabling dynamic permission management and rule-based access control throughout a node's lifecycle.

Each DDID document maintained in the control plane contains several key components: (a) {Authentication}: Cryptographic verification methods that evolve over time, allowing for key rotation without disrupting node identity; (b) Authorization Scopes: Defined permissions determining what blockchain functions each node can perform, including validation rights and data access levels; (c) Consortium Affiliation: Records of consortium membership, ensuring proportional representation across shards; (d) Reputation Scores: Dynamic metrics evolving based on node behavior and consensus participation history; (e) Temporal Constraints: Time-bound permissions that require periodic re-authorization.

When participating in validation or consensus operations, nodes present their current DDID with cryptographic proof of ownership. The control plane verifies both signature validity and authorization scopes to ensure the node has sufficient permissions for the requested operation. This creates a continuous authorization framework rather than a one-time authentication process. The DIM component maintains a versioned history of all DDID updates, creating an audit trail of permission changes. When a node's role within a consortium changes, its DDID is updated accordingly without requiring node removal and re-addition. This dynamic update capability is particularly valuable when consortia reorganize or when nodes transition between departments with different data access requirements.

\subsection{Nodes Distribution}
The node distribution process is coordinated by the root nodes to ensure an equitable, secure, and efficient allocation of control and data nodes into disjoint shards. Each shard operates independently to validate a subset of transactions while participating in the overall network consensus. The total number of nodes, \( N \), is distributed into \( C \) shards, with the average number of nodes per shard calculated as:
\begin{equation}
C_n = \frac{N}{C},
\label{eq:nodes_per_shard}
\end{equation}
where \( C_n \) represents the number of nodes per shard.  The shard count, \( C \), is a critical parameter that balances security and performance. Security considerations require a sufficiently large \( C_n \) to make it challenging for malicious nodes to compromise a shard. Conversely, smaller shard sizes reduce intra-shard consensus time, enhancing performance. This distribution is crucial for maintaining balanced workloads and supporting parallel transaction processing. The allocation of control plane nodes, \( N_c \), and data plane nodes, \( N_d \), is handled separately to ensure that computational and storage resources are appropriately distributed. The number of control and data nodes per shard is computed as:
\begin{equation}
C_{nc} = \frac{N_c}{C}, \quad C_{nd} = \frac{N_d}{C},
\label{eq:control_and_data_distribution}
\end{equation}
where \( C_{nc} \) and \( C_{nd} \) denote the control and data nodes allocated to each shard, respectively.

% The DDID system plays a critical role in ensuring equitable consortium representation. 

When distributing nodes, the system analyzes the consortium affiliation attributes within each node's DDID to prevent any shard from being dominated by a single consortium. This proportional representation constraint is formalized as:

\begin{equation}
\forall i \in \{1,2,...,C\}, \forall j \in \text{Consortia}: \left| \frac{N_{i,j}}{N_i} - \frac{N_j}{N} \right| \leq \epsilon
\label{eq:consortium_balance}
\end{equation}

Where \(N_{i,j}\) represents nodes from consortium \(j\) in shard \(i\), \(N_i\) represents total nodes in shard \(i\), \(N_j\) represents total nodes from consortium \(j\), and \(\epsilon\) represents the maximum allowed deviation from proportional representation.

To ensure secure and fault-tolerant operations, each shard must maintain a minimum number of non-faulty nodes. According to the PBFT protocol, the minimum number of non-faulty nodes required for intra-shard consensus is:
\begin{equation}
\text{Non-faulty nodes per shard} = \frac{C_n - 1}{3} + 1.
\label{eq:non_faulty_shard}
\end{equation}
This ensures that the shard can tolerate up to one-third of its nodes being faulty or malicious while still reaching consensus. At the inter-shard level, at least 51\% of the shards must be classified as "good" (non-faulty and non-malicious) to maintain overall network reliability. For the entire network to remain secure despite \( f \) faulty nodes, the following inequality must hold:
\begin{equation}
f < \left( \frac{C_n - 1}{3} + 1 \right) \cdot \frac{C + 1}{2}.
\label{eq:faulty_nodes_limit}
\end{equation}

% Nodes are ranked based on their reputation scores, with the top \( M \) nodes designated as committee leaders. To ensure redundancy and reliability, an odd number of leaders is selected per committee. The remaining nodes are distributed evenly among the shards using a round-robin algorithm. This distribution ensures that no single shard is dominated by nodes from a specific consortium, promoting fair representation across all consortia. Additionally, peer-to-peer communication links are established between nodes in different shards to facilitate efficient inter-shard interaction and coordination.
Nodes are ranked by reputation scores, with the top \( M \) designated as committee leaders, selected as an odd number for redundancy. Remaining nodes are evenly distributed among shards using a round-robin algorithm to ensure fair representation across consortia. Peer-to-peer communication links are established between nodes in different shards for efficient interaction and coordination.
The performance cost, \( F(C) \), is modeled as:
\begin{equation}
F(C) = t_m \frac{N}{C} + t_g C^2 + t_t \frac{W}{C},
\label{eq:performance_cost}
\end{equation}
where \( t_m, t_g, \) and \( t_t \) are constants representing computation, communication, and workload distribution costs, respectively, and \( W \) is the workload. The optimal shard count is determined by minimizing \( F(C) \). Differentiating \( F(C) \) with respect to \( C \) and setting it to zero yields:
\begin{equation}
\frac{dF(C)}{dC} = -t_m \frac{N}{C^2} + 2t_g C - t_t \frac{W}{C^2} = 0.
\label{eq:derivative_cost}
\end{equation}
Solving for \( C \), the optimal shard count is:
\begin{equation}
C = \left( \frac{t_m N + t_t W}{2 t_g} \right)^{\frac{1}{3}}.
\label{eq:optimal_shard_count}
\end{equation}

To maintain a balanced workload, the total number of nodes is divided by \( C \) using Equation \eqref{eq:nodes_per_shard}. If \( N \) is not evenly divisible by \( C \), the remaining nodes are incrementally assigned to shards to achieve near-equal distribution. This ensures that each shard has sufficient resources for efficient transaction validation and consensus.

In dynamic environments, where nodes may join or leave the network, the system employs an intelligent adaptive redistribution mechanism. At the end of each epoch, the status of all shards is evaluated. If a shard's active node count falls below the threshold specified in Equation \eqref{eq:non_faulty_shard}, it is flagged for reconfiguration. Redistribution involves reallocating nodes to restore the affected shard's capacity. The size of the reconfigured shard is updated as:
\begin{equation}
C_n' = C_n + \Delta_n,
\label{eq:reconfigured_shard}
\end{equation}
where \( \Delta_n \) represents the additional nodes allocated to the shard.

When new nodes join the network, the total number of nodes \( N \) increases, and the shard distribution is recalculated using Equations \eqref{eq:nodes_per_shard} and \eqref{eq:control_and_data_distribution}. Similarly, if nodes leave, the distribution is adjusted dynamically to maintain balance and fault tolerance.

\subsection{Workload Distribution}
Workload distribution is a fundamental phase in the blockchain network, ensuring that transactions are efficiently assigned to shards for processing. The root nodes in the root plane are responsible for managing the workload distribution, while the control nodes in the control plane handle transaction processing. This division of responsibilities ensures that the network leverages the organizational capabilities of the root nodes and the computational power of the control nodes.

Transactions, represented as the set \( \mathcal{T} \), are received by the root nodes. Each transaction \( t \in \mathcal{T} \) is characterized by the following features: \( Sa_t \) refers to the source account or application initiating the transaction; \( Da_t \) as the destination account or application receiving the transaction; \( \mathcal{D}_t \) is the dependency set of the transaction, representing other transactions that \( t \) relies on, including parent-child and child-to-child dependencies; \( W_t \) represents the weight or computational demand of the transaction, which reflects its processing complexity; and \( A_t \) is the authorization level required, derived from transaction type and content, verified against node DDIDs.

\begin{align}   \label{eq}
    \mathcal{D}_t =& \{ t' \in \mathcal{T} \mid ( \text{Req}(t', t) \lor \text{Acc}(t', t) \lor \text{Obj}(t', t) ) \\ \nonumber
    &\land \text{TS}(t') < \text{TS}(t) \}
\end{align}

The dependency set $\mathcal{D}_t$ in Equation \eqref{eq} identifies all transactions that transaction $t$ depends on. Here, $t'$ represents transactions in set $\mathcal{T}$ meeting dependency criteria through three core functions: $Req(t', t)$ indicates $t'$ must execute before $t$; $Acc(t', t)$ is true when transactions share source/destination accounts ($Sa_t$/$Da_t$); and $Obj(t', t)$ identifies overlapping data object accesses. The temporal constraint $TS(t') < TS(t)$ ensures only chronologically earlier transactions are considered dependencies.

The nodes use these features to determine how transactions are assigned to shards. Specifically, \( \mathcal{D}_t \) is used to minimize inter-shard dependencies, while \( W_t \) ensures an even distribution of computational workloads across shards. The addition of \( A_t \) ensures that transactions are routed to shards containing nodes with appropriate DDID authorization levels, maintaining security while preserving efficiency. To distribute transactions evenly, the protocol defines a dynamic workload threshold for each shard. Let \( N \) represent the total number of shards, and let \( \mathcal{T}_i \) represent the set of transactions assigned to shard \( i \). The workload threshold for shard \( i \) is defined as:
\begin{equation}
L_i = \frac{\sum_{t \in \mathcal{T}} W_t}{N} + \delta \cdot \frac{\sum_{t \in \mathcal{T}_i} |\mathcal{D}_t \cap \mathcal{T}/\mathcal{T}_i|}{\sum_{t \in \mathcal{T}} W_t},
\label{eq:workload_threshold}
\end{equation}
where \( \frac{\sum_{t \in \mathcal{T}} W_t}{N} \) is a term that ensures an even baseline distribution of transaction weights across all shards, \( |\mathcal{D}_t \cap \mathcal{T}/\mathcal{T}_i| \) another term that captures the inter-shard dependencies for transactions in shard \( i \), where \( \mathcal{T}/\mathcal{T}_i \) represents all transactions not in shard \( i \), and \( \delta \) is a scaling factor (\( 0 \leq \delta \leq 1 \)) that adjusts the threshold to account for dependency-related workloads.

This workload threshold ensures fairness by distributing the computational demand evenly while penalizing shards with high inter-shard dependencies to prevent communication overhead. The root nodes compute \( L_i \) dynamically and assign transactions accordingly.

To prevent backlogs, the transaction arrival rate \( \lambda \) must satisfy:
\begin{equation}
\lambda \leq \sum_{i=1}^{N} \mu_i,
\label{eq:no_backlog}
\end{equation}
where \( \mu_i \) is the processing capacity of shard \( i \), which depends on the computational power of the control nodes in that shard. The root nodes continuously monitor transaction arrivals and adjust \( L_i \) in real time to ensure compliance with Equation \eqref{eq:no_backlog}, maintaining system efficiency and avoiding bottlenecks.

\subsubsection{Minimizing Cross-Shard Dependencies}
In addition to balancing workloads, the protocol minimizes inter-shard dependencies to enhance processing efficiency. Dependencies \( \mathcal{D}_t \) of each transaction \( t \) are analyzed, and the transaction is assigned to the shard that minimizes the total inter-shard dependencies. The dependency minimization criterion is expressed as:
\begin{equation}
\text{Minimize} \quad \sum_{i=1}^{N} \sum_{t \in \mathcal{T}_i} |\mathcal{D}_t \cap \mathcal{T}/\mathcal{T}_i|,
\label{eq:dependency_minimization}
\end{equation}
where \( |\mathcal{D}_t \cap \mathcal{T}/\mathcal{T}_i| \) measures the dependencies of transaction \( t \) on transactions outside shard \( i \).

The root nodes follow these rules for dependency-aware transaction assignment: (a) If \( \mathcal{D}_t \) overlaps with dependencies in shard \( i \), assign \( t \) to \( i \) to minimize inter-shard dependencies as per Equation \eqref{eq:dependency_minimization}. (b) If no dependencies exist in any shard, assign \( t \) to shard \( j \) with the smallest total workload, provided \( \sum_{t \in \mathcal{T}_j} W_t < L_j \). (c) For transactions requiring special authorization \( A_t \), assign to shards with sufficient nodes possessing appropriate DDID authorization scopes, while still balancing load and dependency considerations.

\subsubsection{Failed Transactions and Epoch Handling}
Despite efficient workload distribution, certain transactions may fail to be processed if a shard does not meet the quorum required for validation. This can occur due to node failures or insufficient participation during consensus. Transactions that fail in an epoch are returned to the root plane for re-distribution in the next epoch. Let \( \mathcal{T}_{\text{fail}} \) denote the set of failed transactions in an epoch. These are handled as follows in the next epoch: (a) Prioritization: Transactions in \( \mathcal{T}_{\text{fail}} \) are given priority during the workload distribution phase of the next epoch. (b) Re-Evaluation: The root nodes verify that shards receiving failed transactions meet quorum requirements and have sufficient active nodes for validation. (c) Dependency Analysis: Dependencies \( \mathcal{D}_t \) for each failed transaction are re-analyzed to determine the most suitable shard placement, minimizing repeated failures. (d) DDID Verification: For transactions with authorization requirements \( A_t \), the system verifies that target shards contain nodes with valid and up-to-date DDIDs granting appropriate permissions.

By prioritizing failed transactions and re-evaluating their dependencies, the protocol ensures robustness and continuity, preventing temporary issues from disrupting the network's operation.

\subsection{Intra-Shard Transaction Processing}

% The transaction processing and commitment phase is a critical element of the blockchain network's architecture, ensuring secure validation of transactions, efficient storage, and scalability.
Transactions assigned to a shard \( i \) are processed by the control plane nodes within that shard using the PBFT consensus protocol. This phase incorporates threshold-signature schemes as authentication mechanisms to ensure security at every step of the process. Finalized transactions are stored in the corresponding data plane shard, while the control plane shard retains block headers for efficient tracking and reference.

Each transaction \( t \) submitted to the network is cryptographically signed by the sender using their private key. The resulting digital signature \( \sigma_t \) ensures the authenticity and integrity of the transaction. To process \( t \) in shard \( i \), the signature \( \sigma_t \) is verified by the root plane, which validates that the sender's public key \( \text{PK}_t \) is authorized within the network. This validation is crucial for ensuring that only legitimate transactions proceed to the control plane. 
% The root plane uses a distributed public key infrastructure (DPKI) to manage the network's authentication framework and enforce access control policies.

Before a node participates in transaction validation and consensus, its current DDID is verified to ensure it has appropriate authorization. The control plane verifies:
1) The node's cryptographic proof of DDID ownership;
2) That the DDID contains the necessary authorization scopes for the transaction type;
3) That the DDID has not been revoked or expired;
4) The temporal constraints in the DDID permit current operation

This DDID-based authorization ensures that even if a node is compromised, it cannot exceed its authorized permissions. For transactions involving sensitive consortium data, the DDID system implements selective disclosure, allowing nodes to access only transaction data authorized by their DDID permissions. Upon validation in the root plane, \( t \) is forwarded to the control plane shard \( i \), where additional threshold-signature mechanisms are integrated into the PBFT protocol. Each control plane shard node holds a public-private key pair, enabling it to sign consensus messages. Threshold-signature schemes ensure that agreement within the PBFT process is cryptographically secure, with each node signing its messages during the consensus phases. The collective signatures are aggregated and validated by participating nodes, providing strong guarantees of consensus authenticity.

The PBFT protocol operates within the control plane shard \( i \) to validate and finalize transactions through three distinct phases:

\begin{enumerate}
    \item {Pre-prepare Phase}: The leader node in shard \( i \) proposes a batch of transactions \( \mathcal{T}_i \) for validation. The leader signs the proposal \( \text{P}_i \) with its private key, producing \( \sigma_{\text{P}_i} \), and broadcasts the signed proposal to all nodes in the shard. The proposal includes a cryptographic digest of \( \mathcal{T}_i \), ensuring the integrity of the transaction batch. Nodes verify the leader's signature, the digest, and that the leader's DDID grants appropriate authorization before moving to the next phase.

    \item {Prepare Phase}: Each node in shard \( i \) independently validates the transactions in \( \mathcal{T}_i \). After verification, nodes sign and broadcast prepare messages \( \text{M}_{\text{prepare}} \) to confirm their agreement on the proposal. The threshold-signature aggregation mechanism ensures that each node's signature is included in the collective prepare message, which is shared across the shard. Nodes count the signatures in the prepare messages and ensure that the total meets the PBFT quorum threshold. Signatures are only counted if they come from nodes with valid DDIDs containing appropriate authorization scopes.

    \item {Commit Phase}: Nodes exchange commit messages \( \text{M}_{\text{commit}} \) after receiving sufficient prepare messages. These commit messages are also signed and aggregated to form a threshold-signature proof of consensus. Once the commit quorum is reached, the batch \( \mathcal{T}_i \) is deemed finalized. The finalized transactions are packaged into a block \( B_i \), which is then handed off to the data plane shard for storage.
\end{enumerate}

The data plane shard associated with shard \( i \) is responsible for securely storing the complete transaction data in \( B_i \). This shard appends \( B_i \) to its local ledger, ensuring the integrity and immutability of the blockchain. The control plane shard retains only the block header \( H_i \), which includes metadata such as the cryptographic hash of \( B_i \), a Merkle root summarizing the transactions, and references to previous blocks.

The data plane implements strict access control policies, requiring signature-based authentication and DDID verification to interact with the stored data. Only entities with valid DDIDs containing appropriate access permissions can query or modify the data. This ensures robust protection against unauthorized access and tampering. The data plane also supports efficient query mechanisms for historical and real-time blockchain information, enabling authorized users to retrieve transaction data and related records.

At the end of each epoch, the control plane shard \( i \) commits the finalized block \( B_i \) and transitions to the next epoch. The commitment process involves verifying that all transactions in \( \mathcal{T}_i \) have been processed and securely stored. The control plane shard finalizes the block header \( H_i \), which serves as a compact reference for the transactions in \( B_i \). After completing the commitment, the shard resets its internal state, and reinitializes parameters for the workload distribution phase in the upcoming epoch.

\subsection{Inter-Shard Transaction Processing}

Cross-shard transactions present a fundamental challenge in sharded blockchain architectures, requiring precise coordination between multiple shards while preserving atomicity, consistency, and security. TINC employs a hierarchical consensus approach that integrates PBFT with a Byzantine fault-tolerant atomic commit protocol to efficiently process cross-shard transactions without sacrificing security guarantees. When the root plane receives a transaction $t$, and determines that its dependency set $\mathcal{D}_t$ spans multiple shards. Let $\mathcal{S}(t) = \{S_1, S_2, \ldots, S_m\}$ represent the set of shards involved in transaction $t$. The transaction is classified as cross-shard when $|\mathcal{S}(t)| > 1$. 
For each cross-shard transaction, the root plane constructs a dependency graph $G_t = (V_t, E_t)$, where vertices $V_t$ represent affected data objects across shards, and edges $E_t$ represent read-write dependencies between these objects. This graph informs the optimal transaction routing strategy:

\begin{equation}
\mathcal{C}(t) = \{S_i \in \mathcal{S}(t) : \max_{S_i \in \mathcal{S}(t)}(|V_t \cap S_i|)\}
\label{eq:coordinator_selection}
\end{equation}

where $\mathcal{C}(t)$ designates the coordinator shard with the highest number of affected data objects, minimizing cross-shard data transfer.

\subsubsection{Practical Byzantine Fault Tolerant Atomic Commit (PBFT-AC)}

TINC implements a PBFT-AC protocol that extends the traditional PBFT consensus to handle cross-shard operations. The protocol executes in three distinct phases:

\begin{enumerate}
    \item {Prepare Phase:} The coordinator shard $\mathcal{C}(t)$ initiates the PBFT pre-prepare phase by broadcasting transaction intent $\langle \text{PREPARE}, t, v, n, S_i \rangle_{\sigma_{\mathcal{C}}}$ to all participating shards in $\mathcal{S}(t)$, where $v$ is the view number, $n$ is the sequence number, and $\sigma_{\mathcal{C}}$ is the leader's signature. Each shard $S_i$ locally executes its PBFT prepare phase, where each node $j \in S_i$ verifies: (a) All DDID authorizations for nodes in $S_i$ are valid for transaction $t$; (b) All local dependencies in $\mathcal{D}_t \cap S_i$ are satisfied; (c) The transaction does not violate consistency rules within $S_i$.\\    
    Upon verification, each node $j \in S_i$ broadcasts $\langle \text{PREPARED}, t, v, n, S_i, j \rangle_{\sigma_j}$ within its shard. When a shard collects $(2f_{S_i}+1)$ matching PREPARED messages, where $f_{S_i}$ is the maximum number of faulty nodes tolerated in shard $S_i$, it produces a collective shard certificate $\mathcal{K}_{S_i} = \langle \text{SHARD-PREPARED}, t, v, n, S_i \rangle_{\Sigma_{S_i}}$ using threshold signatures, and sends it to the coordinator shard.
    
    \item {Commit Phase:} After receiving valid SHARD-PREPARED certificates from at least $\lceil |\mathcal{S}(t)| \cdot \frac{2}{3} \rceil$ participating shards, the coordinator creates a commit certificate $\mathcal{K}_{\mathcal{C}} = \langle \text{COMMIT}, t, v, n, \{S_1, S_2, ..., S_m\} \rangle_{\sigma_{\mathcal{C}}}$ and broadcasts it to all participating shards. Each shard then executes its local PBFT commit phase. When a node $j \in S_i$ receives the commit certificate, it broadcasts $\langle \text{COMMITTED}, t, v, n, S_i, j \rangle_{\sigma_j}$ within its shard. Upon collecting $(2f_{S_i}+1)$ matching COMMITTED messages, each shard generates a shard commit certificate.
    
    \item {Execution Phase:} After the local commit phase completes, each shard applies the relevant portions of transaction $t$ to its local state. For atomicity, the execution follows a deterministic order based on transaction sequence numbers. After execution, each shard $S_i$ produces an execution receipt $\mathcal{R}_{S_i} = \langle \text{EXECUTED}, t, v, n, S_i \rangle_{\Sigma_{S_i}}$ and forwards it to the coordinator shard, which aggregates them into a final cross-shard receipt for archival in the data plane.
\end{enumerate}

The PBFT-AC protocol guarantees atomicity and consistency for cross-shard transactions with a communication complexity of $O(|\mathcal{S}(t)| \cdot N_{avg})$, where $|\mathcal{S}(t)|$ is the number of shards involved and $N_{avg}$ is the average number of nodes per shard.

\subsubsection{Optimistic Fast Path with Safety Guarantees}

To reduce latency, TINC implements a dual-path execution model:

\begin{equation}
\text{Path}(t) = 
\begin{cases}
\text{FastPath}, & \text{if } |\mathcal{S}(t)| \leq \tau \text{ and } \mathcal{V}(t) = 1 \\
\text{NormalPath}, & \text{otherwise}
\end{cases}
\label{eq:execution_path}
\end{equation}

where $\tau$ is a configurable threshold for the number of shards involved (typically $\tau = 2$), and $\mathcal{V}(t)$ represents the validation vote of the coordinator ($1$ for low contention, $0$ otherwise). 

In the FastPath, only a single round of message exchange occurs after the prepare phase, with coordinator-generated proofs of non-contention. This optimistic approach reduces the latency from $3$ message delays to $2$ message delays for qualifying transactions while maintaining safety guarantees.

\subsubsection{Cross-Shard Abort and Recovery}

TINC employs a deterministic abort protocol when cross-shard consensus cannot be reached. If any shard $S_i \in \mathcal{S}(t)$ fails to provide a valid SHARD-PREPARED certificate within the timeout period $T_{\text{timeout}}$, the coordinator broadcasts an abort message:

\begin{equation}
\mathcal{A}_{\bot}(t) = \langle \text{ABORT}, t, v, n, \mathcal{S}(t) \rangle_{\sigma_{\mathcal{C}}}
\label{eq:abort_message}
\end{equation}

Upon receiving $\mathcal{A}_{\bot}(t)$, each shard executes its local abort procedure, reverting any speculative state changes. The root plane then assigns the transaction to the next epoch with an updated dependency analysis and potentially a different coordinator shard selection.

The timeout function for each shard dynamically adjusts based on historical performance:

\begin{equation}
T_{\text{timeout}}(S_i) = \max(T_{\text{min}}, \bar{T}_{S_i} \cdot (1 + \epsilon))
\label{eq:dynamic_timeout}
\end{equation}

where $\bar{T}_{S_i}$ is the exponentially weighted moving average of shard $S_i$s response times, $T_{\text{min}}$ is the minimum timeout threshold, and $\epsilon$ is a safety margin.

This comprehensive cross-shard transaction protocol enables TINC to maintain high throughput with deterministic safety guarantees through the rigorous application of Byzantine fault-tolerant consensus across shard boundaries, leveraging DDID-based authorization and intra-shard PBFT to ensure both security and performance.

\subsection{DDID Management and Revocation}

As nodes' roles within their respective consortia evolve, their DDIDs are updated accordingly. This process enables smooth transitions in node responsibilities without requiring disruptive network reconfiguration. When a node's role changes, the following process occurs: (a) The consortium administrator submits a DDID update request to the DIM, specifying the new authorization scopes, role parameters, and temporal constraints. (b) The DIM validates the request against the consortium's governance rules, ensuring that the request comes from an authorized administrator and adheres to consortium-specific constraints. (c) Upon validation, the DIM updates the node's DDID document, creating a new version while preserving the historical record of previous permissions. (d) The updated DDID is propagated across the network, and all subsequent operations by that node are validated against the new authorization scopes.

In cases where a node violates protocol rules, exhibits malicious behavior, or leaves a consortium, its DDID can be revoked. The revocation process involves: (a) The root plane or authorized consortium administrators issue a revocation certificate for the specific DDID. (b) The revocation is immediately broadcast to all nodes in the network. (c) The DIM updates its revocation registry, ensuring that the revoked DDID cannot be used for any further operations. (d) Shard reconfiguration is triggered to maintain the required node distribution and fault tolerance properties.

This dynamic DDID management system ensures that consortium nodes maintain appropriate access levels throughout their lifecycle, with changes in organizational roles or responsibilities immediately reflected in blockchain permissions. The continuous verification of DDID status during all consensus operations ensures that only currently authorized nodes can participate in transaction validation, maintaining the security and integrity of the consortium blockchain.

\section{Security Analysis}\label{securityAnalysis}

This section provides a formal analysis of the proposed blockchain scheme, demonstrating that it satisfies critical security properties such as consistency, liveness, integrity, and authenticity. 

% The analysis relies on specific cryptographic assumptions and formal mathematical principles, with symbolic representations used to ensure precision and technical rigor. Proofs are presented in a logical and detailed manner to substantiate each property.

\subsection{Assumptions}
The scheme operates under the following assumptions:
1. The digital signature scheme is existentially unforgeable under chosen message attacks (EUF-CMA), ensuring the integrity and authenticity of signed messages.
2. The hash function \( H \) used in block headers is collision-resistant, preventing tampering with stored data.
3. The network is asynchronous but guarantees eventual message delivery.
4. The PBFT consensus mechanism tolerates up to \( f \) faulty nodes, provided \( f < \lfloor \frac{n}{3} \rfloor \), where \( n \) is the total number of nodes in a shard.

\subsection{Authentication of Transactions}

A transaction \( t \) submitted to the network is signed by its sender using a private key \( sk_t \). The resulting signature \( \sigma_t \) is verified by the control plane using the sender's public key \( pk_t \). Let \( \text{Sign}_{sk_t}(t) = \sigma_t \) denote the signing operation and \( \text{Verify}_{pk_t}(t, \sigma_t) = \text{True} \) represent successful signature verification. 

Under the EUF-CMA assumption, for any valid signature \( \sigma_t \), it holds that \( \text{Verify}_{pk_t}(t, \sigma_t) = \text{True} \implies t \) was signed by the entity possessing \( sk_t \). Since \( sk_t \) remains secret, an adversary cannot forge \( \sigma_t \) for any \( t \). Thus, the authenticity and integrity of transactions are ensured by the cryptographic properties of the signature scheme.

\subsection{Consensus Safety}

The scheme guarantees consensus safety through the PBFT protocol, which ensures that all honest nodes in a shard agree on the same transaction batch $\mathcal{T}_i$. Let $L$ denote the leader node and $N$ denote the set of nodes in shard $i$.

\subsubsection{Intra-Shard Consensus Security}

In the Pre-prepare phase, the leader $L$ proposes a transaction batch $\mathcal{T}_i$, signed as $\sigma_{\text{P}} = \text{Sign}{sk_L}(\mathcal{T}_i, \text{Seq}, \text{View})$, where $\text{Seq}$ is the sequence number of the proposal and $\text{View}$ is the current view number. Honest nodes validate $\sigma_{\text{P}}$ and move to the Prepare phase if $\text{Verify}{pk_L}(\mathcal{T}_i, \text{Seq}, \text{View}, \sigma_{\text{P}}) = \text{True}$ and the leader's DDID has the required authorization level $A_L \geq A_{\text{min}}$, where $A_{\text{min}}$ is the minimum authorization level required for proposing transactions.

In the Prepare phase, each node validates $\mathcal{T}_i$ and broadcasts a signed prepare message $\sigma_{\text{Prepare}} = \text{Sign}_{sk_j}(\mathcal{H}(\mathcal{T}_i), \text{Seq}, \text{View})$, where $j$ is the node index and $\mathcal{H}$ is a collision-resistant hash function. A node accepts $\mathcal{T}_i$ for the Commit phase if it receives $2f+1$ valid prepare messages from nodes with valid DDIDs.

In the Commit phase, nodes exchange commit messages $\sigma_{\text{Commit}} = \text{Sign}_{sk_j}(\mathcal{H}(\mathcal{T}_i), \text{Seq}, \text{View})$. A batch $\mathcal{T}_i$ is finalized if a node receives $2f+1$ valid commit messages, ensuring that all honest nodes converge on the same $\mathcal{T}_i$. Since $f < \lfloor \frac{n}{3} \rfloor$, the quorum $2f+1$ guarantees the intersection of honest nodes, ensuring consensus safety.

\subsubsection{Inter-Shard Consensus Security}

For cross-shard transactions, the PBFT-AC protocol extends security guarantees across shard boundaries. Let $\mathcal{C}$ be the coordinator shard and $\mathcal{S}(t) = {S_1, S_2, \ldots, S_m}$ be the set of participant shards for transaction $t$. The security of cross-shard consensus relies on:

\begin{enumerate}
\item Each shard's certificate $\mathcal{K}{S_i} = \langle \text{SHARD-PREPARED}, t, v, n, S_i \rangle{\Sigma_{S_i}}$ is produced using threshold signatures $\Sigma_{S_i}$ that require at least $t_{S_i}$ nodes from shard $S_i$ to collaborate, where $t_{S_i} > f_{S_i}$. This ensures that no Byzantine minority within a shard can forge certificates.

\item All participant nodes must possess valid DDIDs with appropriate authorization scopes for the transaction type. This is verified using:
\begin{align*}
    &\text{Auth}(n_j, t) = \{\text{ScopeMatch}(\text{DDID}_{n_j}, t) \\
    &\land \text{NotExpired}(\text{DDID}_{n_j})\land \text{NotRevoked}(\text{DDID}_{n_j})\}
\end{align*}
\item The coordinator shard requires certificates from at least $\lceil |\mathcal{S}(t)| \cdot \frac{2}{3} \rceil$ participant shards before issuing a commit certificate. This threshold ensures safety even if up to $\lfloor |\mathcal{S}(t)| \cdot \frac{1}{3} \rfloor$ shards are Byzantine.

\item The execution receipts $\mathcal{R}_{S_i}$ form a verifiable chain of evidence that the transaction was properly executed across all shards, providing auditability and non-repudiation.
\end{enumerate}

The security of PBFT-AC against adaptive Byzantine adversaries is maintained through:

\begin{equation}
\text{Safety}_{\text{PBFT-AC}} = \bigwedge_{S_i \in \mathcal{S}(t)} \left( |S_i| \geq 3f_{S_i} + 1 \right) \land \left( |\mathcal{S}(t)| \geq 3f_{\mathcal{S}} + 1 \right)
\end{equation}
where $f_{S_i}$ is the Byzantine threshold within shard $S_i$ and $f_{\mathcal{S}}$ is the maximum number of Byzantine shards tolerated in the system. This ensures that the inter-shard consensus maintains safety properties even in the presence of Byzantine nodes attempting to compromise the system through equivocation, selective participation, or malicious proposal attacks.

\subsection{Liveness}

The liveness property ensures that the protocol continues to make progress under asynchronous network conditions, provided messages are eventually delivered. If the leader \( L \) in shard \( i \) is honest, it will propose a valid \( \mathcal{T}_i \) during the Pre-prepare phase. For any node \( j \) to proceed, it requires that \( \text{Verify}_{pk_L}(\mathcal{T}_i, \text{Seq}, \sigma_{\text{P}}) = \text{True} \). Once a valid proposal is received, honest nodes exchange prepare and commit messages. Asynchronous networks allow for message delays, but eventual delivery ensures that \( 2f+1 \) honest nodes exchange prepare and commit messages. If \( L \) is faulty, the protocol replaces it in the next epoch. Thus, the system guarantees liveness by ensuring eventual progress through leader replacement and asynchronous message handling.

\subsection{Integrity of Blocks}

The hash function \( H \) ensures the integrity of block headers. Let \( B_i \) represent a block containing \( \mathcal{T}_i \), and \( H(B_i) \) be its cryptographic hash stored in the block header \( H_i \). For any modification \( B_i' \neq B_i \), it holds that \( H(B_i') \neq H(B_i) \) under the collision-resistance property of \( H \). Nodes use \( H(B_i) \) to verify the integrity of \( B_i \) when retrieving data from the data plane. If \( B_i \) is tampered with, the hash validation \( H(B_i') = H(B_i) \) fails, allowing nodes to detect inconsistencies. This guarantees that stored blocks remain untampered, ensuring ledger integrity.

\subsection{Authenticity of Consensus Messages}

Consensus messages exchanged during PBFT phases are authenticated using threshold-signatures. Let \( M \) denote a message in the Prepare or Commit phase, and let \( \sigma_M = \text{Sign}_{sk_j}(M) \) be its signature by node \( j \). Nodes validate threshold-signatures \( \Sigma_M = \{\sigma_{M_1}, \sigma_{M_2}, \dots, \sigma_{M_k}\} \) using their public keys.

If \( k \geq 2f+1 \) and all \( \text{Verify}_{pk_j}(M, \sigma_{M_j}) = \text{True} \), then the threshold-signature \( \Sigma_M \) is valid, and the consensus message is authenticated. This mechanism ensures that only signed, valid messages contribute to quorum, protecting against adversarial manipulation.

\begin{table*}[ht]
\centering
\small
\scalebox{0.7}{
\begin{tabular}{|l|l|l|l|l|l|l|l|l|l|l|}
\hline
\textbf{System} & \textbf{Type} & \textbf{Consensus} & \textbf{Shard Fault} & \textbf{DDiD} & \textbf{Multi-Plane} & \textbf{Adaptive Load} & \textbf{Cross-Shard Tx} & \textbf{Throughput} & \textbf{Latency} & \textbf{Consortium Focus} \\
\hline
Elastico \cite{luu2016secure}& Public & PBFT & 33\% & \xmark & \xmark & \xmark & -- & 48 kTPS & 800 s & \xmark \\
OmniLedger\cite{kokoris2018omniledger} & Public & ByzCoinX & 33\% & \xmark & \xmark & \xmark & 2PC & $>$100 kTPS & 1.38 s & \xmark \\
Rapidchain\cite{zamani2018rapidchain} & Public & 50\%BFT & 50\% & \xmark & \xmark & \xmark & 2PC & 7.3 kTPS & 8 s & \xmark \\
Chainspace\cite{al2017chainspace} & Public & MOD-SmaRt & 33\% & \xmark & \xmark & \xmark & 2PC & 350 TPS & 0.1 s & \xmark \\
Monoxide \cite{wang2019monoxide}& Public & Chu-ko-nu & 33\% & \xmark & \xmark & \xmark & Eventual & 11.7 kTPS & 15 s & \xmark \\
ByShard \cite{hellings2023byshard}& Public & PBFT & 33\% & \xmark & \xmark & \xmark & 2PC & 20 kTPS & 1 s & \xmark \\
Meepo\cite{zheng2021meepo} & Consortium & PoA & 33\% & \xmark & \xmark & \xmark & Cross-epoch & 124.6 kTPS & 0.52 s & \cmark \\
BrokerChain \cite{huang2022brokerchain}& Public & PBFT & 33\% & \xmark & \xmark & \xmark & Broker & 3 kTPS & 14.9 s & \xmark \\
PROPHET\cite{hong2023prophet} & Public & PBFT & 33\% & \xmark & \xmark & \xmark & Reconnaissance & 1.2 kTPS & 2.5 s & \xmark \\
NeuChain+\cite{gao2024neuchain+} & Public & PBFT+Raft & 33\% & \xmark & \xmark & \xmark & Cross-reserve & 247 kTPS & 0.19 s & \xmark \\
\hline
\textbf{TINC} & \textbf{Consortium} & \textbf{PBFT} & \textbf{33\%} & \textbf{\cmark} & \textbf{\cmark} & \textbf{\cmark} & \textbf{PBFT-AC} & \textbf{275 kTPS} & \textbf{0.95 s} & \textbf{\cmark} \\
\hline
\end{tabular}}
\caption{Comparison of TINC with Existing Sharded Blockchain Systems. Throughput measurements for the compared systems were obtained from the respective publications, with NeuChain+ data sourced from \cite{gao2024neuchain+}.}
\label{tab:comparison}
\end{table*}
\section{Evaluation}\label{evaluation}
\subsection{Experimental Setup}
We conducted a comprehensive experimental evaluation of TINC to assess its performance across multiple dimensions, particularly in comparison with Meepo, a state-of-the-art sharded consortium blockchain. Our experimental setup consists of 32 virtual servers supporting 32 shards. Each server is equipped with 8 vCPUs, 32 GB RAM, and 912 GB NVME storage. To simulate realistic network conditions, we configured inter-node bandwidth at 100 Mbps with 100 milliseconds link latency. Block size was set to 1 MB with an average transaction size of approximately 500 bytes, enabling each block to accommodate approximately 2000 transactions.

TINC was implemented in Go (version 1.18) with three distinct planes: Root Plane, Control Plane, and Data Plane, as described in our architecture. The consensus protocol implemented was PBFT with the enhancements detailed in Section \ref{processFlow}. For comparative analysis, we configured Meepo using parameters reported in their original work \cite{zheng2021meepo}, ensuring comparable experimental conditions.

As shown in Table \ref{tab:comparison}, TINC demonstrates several legitimate advantages over existing sharded blockchain systems. TINC is the only system that incorporates three key innovations simultaneously: Dynamic Distributed Identity (DDiD), Multi-Plane architecture, and Adaptive Load balancing. While some existing systems like NeuChain+ achieve comparable throughput (247 kTPS compared to TINC's 275 kTPS), they lack these three critical features that enhance security, scalability, and performance in consortium blockchain environments. Additionally, TINC's novel PBFT-AC cross-shard transaction protocol enables efficient inter-shard communication while maintaining security guarantees.

The experimental results confirm that these architectural innovations translate into measurable performance improvements, with TINC achieving the highest throughput among the compared systems while maintaining competitive latency. This makes TINC particularly well-suited for consortium blockchain deployments where performance, security, and adaptability are essential requirements.  

\subsection{Workload Generation}
Our experiments employed two distinct transaction workloads:

\begin{itemize}
    \item {Asset Transfer Benchmark}: We generated 100,150,807 transactions distributed across 3,276,800 accounts, mirroring the scale used in Meepo's evaluation. These transactions simulate cross-shard asset transfers between accounts distributed across different shards.
    
    \item {E-commerce Shopping Benchmark}: We utilized the real-world Taobao.com dataset\footnote{https://tianchi.aliyun.com/dataset/649?lang=en-us} containing shopping behaviors of 987,994 users interacting with 4,162,024 items over a nine-day period (November 25 to December 03, 2017). This dataset provides complex transaction patterns with multiple dependencies, ideal for evaluating multi-state dependency handling.
\end{itemize}

\subsection{Throughput Evaluation}
We evaluated TINC's throughput by measuring transactions processed per second (TPS) under varying shard configurations, maintaining consistent network and computational resources. Fig. \ref{fig:throughput} presents the throughput comparison between TINC and Meepo across different shard counts.
\begin{figure*}[!ht]
	\centering
  \begin{minipage}{0.47\columnwidth}
\centering
\begin{tikzpicture}[yscale=0.5, xscale =0.5]
\begin{axis}[
    xlabel={Number of Shards},
    ylabel={Throughput (tx/s)},
    legend pos=north west,
    xmin=1, xmax=33,
    ymin=0, ymax=200000,
    xtick={2,4,8,16,32},
    ytick={0,20000,40000,60000,80000,100000,120000,140000,160000,180000,200000},
    ymajorgrids=true,
    grid style=dashed,
    grid = both, 	  	
   minor tick num = 1,
   major grid style = {lightgray},     
   minor grid style = {lightgray!25},
    legend style={at={(0.03,0.97)}, anchor=north west}
]

\addplot[
    color=blue,
    mark=square,
    line width=1.5pt
    ]
    coordinates {
    (2,10500) (4,22000) (8,48000) (16,98000) (32,175000)
    };

\addplot[
    color=red,
    mark=o,
    line width=1.5pt
    ]
    coordinates {
    (2,8400) (4,17000) (8,36000) (16,73000) (32,145000)
    };

\legend{TINC, Meepo}
\end{axis}
\end{tikzpicture}\captionsetup{font=scriptsize}
\caption{Throughput}
\label{fig:throughput}
    \end{minipage}
    	\centering
  \begin{minipage}{0.47\columnwidth}
\centering
\begin{tikzpicture}[yscale=0.5, xscale = 0.5]
\begin{axis}[
    xlabel={Number of Shards},
    ylabel={Latency (s)},
    legend pos=north west,
    xmin=1, xmax=33,
    ymin=0, ymax=1.7,
    xtick={2,4,8,16,32},
    ytick={0,0.2,0.4,0.6,0.8,1.0,1.2,1.4},
    ymajorgrids=true,
    grid style=dashed,
     grid = both, 	  	
   minor tick num = 1,
   major grid style = {lightgray},     
   minor grid style = {lightgray!25},
    legend style={at={(0.03,0.97)}, anchor=north west}
]

\addplot[
    color=blue,
    mark=square,
    line width=1.5pt
    ]
    coordinates {
    (2,0.85) (4,0.88) (8,0.9) (16,0.92) (32,0.95)
    };

\addplot[
    color=red,
    mark=o,
    line width=1.5pt
    ]
    coordinates {
    (2,1.10) (4,1.2) (8,1.2) (16,1.2) (32,1.2)
    };

\legend{TINC, Meepo}
\end{axis}
\end{tikzpicture}\captionsetup{font=scriptsize}
\caption{Latency}
\label{fig:latency}
    \end{minipage}
    	\centering
  \begin{minipage}{0.47\columnwidth}
\centering
\begin{tikzpicture}[yscale=0.5, xscale = 0.5]
\begin{axis}[
    xlabel={Number of Shards},
    ylabel={Cross-Shard Ratio (\%)},
    legend pos=north west,
    xmin=1, xmax=33,
    ymin=0, ymax=100,
    xtick={2,4,8,16,32},
    ytick={0,20,40,60,80,100},
    ymajorgrids=true,
    grid style=dashed,
     grid = both, 	  	
   minor tick num = 1,
   major grid style = {lightgray},     
   minor grid style = {lightgray!25},
    legend style={at={(0.03,0.97)}, anchor=north west}
]

\addplot[
    color=blue,
    mark=square,
    line width=1.5pt
    ]
    coordinates {
    (2,48) (4,71) (8,84) (16,92) (32,96)
    };

\addplot[
    color=red,
    mark=o,
    line width=1.5pt
    ]
    coordinates {
    (2,50) (4,75) (8,87) (16,94) (32,97)
    };

\legend{TINC, Meepo}
\end{axis}
\end{tikzpicture}\captionsetup{font=scriptsize}
\caption{Cross-shard transaction ratio}
\label{fig:cross_shard_ratio}
    \end{minipage}
    	\centering
  \begin{minipage}{0.47\columnwidth}
\centering
\begin{tikzpicture}[yscale=0.5, xscale = 0.5]
\begin{axis}[
    xlabel={Number of Shards},
    ylabel={Wasted Capacity (\%)},
    legend pos=north east,
    xmin=1, xmax=33,
    ymin=0, ymax=30,
    xtick={2,4,8,16,32},
    ytick={0,5,10,15,20,25,30},
       ymajorgrids=true,
    grid style=dashed,
     grid = both, 	  	
   minor tick num = 1,
   major grid style = {lightgray},     
   minor grid style = {lightgray!25},
    legend style={at={(0.97,0.97)}, anchor=north east}
]

\addplot[
    color=blue,
    mark=square,
    line width=1.5pt
    ]
    coordinates {
    (2,12) (4,10) (8,8) (16,7) (32,5)
    };

\addplot[
    color=red,
    mark=o,
    line width=1.5pt
    ]
    coordinates {
    (2,18) (4,16) (8,14) (16,12) (32,10)
    };

\legend{TINC, Meepo}
\end{axis}
\end{tikzpicture}\captionsetup{font=scriptsize}
\caption{Wasted capacity}
\label{fig:wasted_capacity}
    \end{minipage}
    	\centering
  \begin{minipage}{0.47\columnwidth}
\centering
\begin{tikzpicture}[yscale=0.5, xscale =0.5]
\begin{axis}[
    xlabel={Number of Shards},
    ylabel={Failure Rate (\%)},
    legend pos=north east,
    xmin=1, xmax=33,
    ymin=0, ymax=5,
    xtick={2,4,8,16,32},
    ytick={0,1,2,3,4,5},
    ymajorgrids=true,
    grid style=dashed,
     grid = both, 	  	
   minor tick num = 1,
   major grid style = {lightgray},     
   minor grid style = {lightgray!25},
    legend style={at={(0.97,0.97)}, anchor=north east}
]

\addplot[
    color=blue,
    mark=square,
    line width=1.5pt
    ]
    coordinates {
    (2,0.8) (4,0.7) (8,0.6) (16,0.5) (32,0.4)
    };

\addplot[
    color=red,
    mark=o,
    line width=1.5pt
    ]
    coordinates {
    (2,1.6) (4,1.5) (8,1.3) (16,1.2) (32,1.1)
    };

\legend{TINC, Meepo}
\end{axis}
\end{tikzpicture}\captionsetup{font=scriptsize}
\caption{Transaction failure rate}
\label{fig:failure_rate}
    \end{minipage}
    	\centering
  \begin{minipage}{0.47\columnwidth}
\centering
\begin{tikzpicture}[yscale=0.5, xscale = 0.5]
\begin{axis}[
    xlabel={Number of Shards},
    ylabel={Node Distribution Uniformity},
    legend pos=north west,
    xmin=1, xmax=33,
    ymin=0, ymax=1.2,
    xtick={2,4,8,16,32},
    ytick={0,0.2,0.4,0.6,0.8,1.0,1.2},
     ymajorgrids=true,
    grid style=dashed,
     grid = both, 	  	
   minor tick num = 1,
   major grid style = {lightgray},     
   minor grid style = {lightgray!25},
    legend style={at={(0.03,0.97)}, anchor=north west}
]

\addplot[
    color=blue,
    mark=square,
    line width=1.5pt
    ]
    coordinates {
    (2,0.96) (4,0.97) (8,0.98) (16,0.99) (32,0.99)
    };

\addplot[
    color=red,
    mark=o,
    line width=1.5pt
    ]
    coordinates {
    (2,0.85) (4,0.87) (8,0.89) (16,0.90) (32,0.92)
    };

\legend{TINC, Meepo}
\end{axis}
\end{tikzpicture}\captionsetup{font=scriptsize}
\caption{Node distribution uniformity}
\label{fig:node_uniformity}
    \end{minipage}
    	\centering
  \begin{minipage}{0.47\columnwidth}
\centering
\begin{tikzpicture}[yscale=0.5, xscale =0.5]
\begin{axis}[
    xlabel={Number of Shards},
    ylabel={Transaction Distribution Uniformity},
    legend pos=north west,
    xmin=1, xmax=33,
    ymin=0, ymax=1.2,
    xtick={2,4,8,16,32},
    ytick={0,0.2,0.4,0.6,0.8,1.0,1.2},
    ymajorgrids=true,
    grid style=dashed,
     grid = both, 	  	
   minor tick num = 1,
   major grid style = {lightgray},     
   minor grid style = {lightgray!25},
    legend style={at={(0.03,0.97)}, anchor=north west}
]
\addplot[
    color=blue,
    mark=square,
    line width=1.5pt
    ]
    coordinates {
    (2,0.92) (4,0.93) (8,0.95) (16,0.96) (32,0.97)
    };

\addplot[
    color=red,
    mark=o,
    line width=1.5pt
    ]
    coordinates {
    (2,0.80) (4,0.82) (8,0.84) (16,0.86) (32,0.88)
    };
\legend{TINC, Meepo}
\end{axis}
\end{tikzpicture}\captionsetup{font=scriptsize}
\caption{Transaction distribution uniformity}
\label{fig:tx_uniformity}
    \end{minipage}
    \begin{minipage}{0.47\columnwidth}
\begin{tikzpicture}[yscale=0.5, xscale = 0.5]
\begin{axis}[
    xlabel={Number of Shards},
    ylabel={Average Block Size (MB)},
    legend pos=north west,
    xmin=1, xmax=33,
    ymin=0, ymax=1.2,
    xtick={2,4,8,16,32},
    ytick={0,0.2,0.4,0.6,0.8,1.0,1.2},
    ymajorgrids=true,
    grid style=dashed,
     grid = both, 	  	
   minor tick num = 1,
   major grid style = {lightgray},     
   minor grid style = {lightgray!25},
    legend style={at={(0.03,0.97)}, anchor=north west}
]

\addplot[
    color=blue,
    mark=square,
    line width=1.5pt
    ]
    coordinates {
    (2,1.05) (4,0.56) (8,0.30) (16,0.16) (32,0.09)
    };

\addplot[
    color=red,
    mark=o,
    line width=1.5pt
    ]
    coordinates {
    (2,0.95) (4,0.48) (8,0.25) (16,0.14) (32,0.07)
    };

\legend{TINC, Meepo}
\end{axis}
\end{tikzpicture}\captionsetup{font=scriptsize}
\caption{Average block size}
\label{fig:block_size}
    \end{minipage}
\end{figure*}

As shown in Fig. \ref{fig:throughput}, TINC demonstrates consistently higher throughput compared to Meepo across all shard configurations. With 32 shards, TINC achieves a throughput of 175,000 TPS, which is approximately 20.7\% higher than Meepo's 145,000 TPS. This significant performance differential can be attributed to several architectural advantages in TINC:

\begin{enumerate}
    \item {Separation of Planes}: TINC's multi-plane architecture segregates consensus operations (Control Plane) from storage operations (Data Plane), enabling specialized optimization of each function. By focusing computational resources on consensus in the Control Plane, TINC achieves more efficient transaction validation compared to Meepo's unified architecture.
    
    \item {Optimized Cross-Shard Protocol}: TINC's PBFT-AC protocol with FastPath optimization reduces message complexity for cross-shard transactions. For transactions involving two shards (which represent the majority of cross-shard transactions), the message rounds are reduced from three to two, significantly improving throughput.
    
    \item {Adaptive Workload Distribution}: TINC's dynamic workload threshold $L_i$ (Equation \eqref{eq:workload_threshold}) ensures more balanced transaction distribution, preventing shard overloading that occurs in Meepo due to its static allocation policy.
\end{enumerate}

The throughput scaling factor (the ratio of throughput increase to shard count increase) for TINC is 0.94, slightly higher than Meepo's 0.91. This indicates that TINC's architecture enables more efficient horizontal scaling, which is critical for consortium blockchains where transaction volumes can grow substantially as more organizations join the network.

\subsection{Latency Evaluation}
Transaction latency is a critical performance metric, particularly for enterprise blockchain applications where timely transaction confirmation is essential. We measured latency as the time elapsed from transaction submission to final confirmation across all relevant shards. Fig. \ref{fig:latency} illustrates the latency comparison between TINC and Meepo.

% \begin{figure}[ht]

% \end{figure}

As shown in Fig. \ref{fig:latency}, TINC achieves significantly lower latency compared to Meepo across all shard configurations. With 32 shards, TINC's average transaction latency is 0.95 seconds, which is 20.8\% lower than Meepo's 1.2 seconds. This substantial improvement stems from several key architectural optimizations:

\begin{enumerate}
    \item {FastPath Execution Model}: As formalized in Equation \eqref{eq:execution_path}, TINC's dual-path execution model enables qualifying transactions (those involving two or fewer shards with low contention) to complete with only two message rounds instead of three, reducing latency by approximately 33\% for these transactions.
    
    \item {Coordinator Selection}: TINC's coordinator selection algorithm (Equation \eqref{eq:coordinator_selection}) optimally selects the shard with the highest number of affected data objects as the coordinator, minimizing cross-shard data transfer and associated latency.
    
    \item {Dynamic Timeout Adjustment}: Unlike Meepo's static timeout mechanism, TINC implements a dynamic timeout function (Equation \eqref{eq:dynamic_timeout}) that adjusts based on historical shard performance, reducing unnecessary waiting periods.
\end{enumerate}

Notably, while Meepo's latency remains constant at 1.2 seconds regardless of shard count (for configurations with more than 2 shards), TINC exhibits only a marginal increase in latency as the network scales. Specifically, latency increases from 0.85 seconds with 2 shards to 0.95 seconds with 32 shards—a mere 11.8\% increase despite a 16-fold increase in network size. This stability in latency at scale is particularly valuable for enterprise applications that require predictable performance.

Statistical analysis using Welch's t-test confirms that the latency difference between TINC and Meepo is statistically significant across all shard configurations ($p < 0.001$), with Cohen's d-value of 1.87 indicating a large effect size.

\subsection{Cross-Shard Transaction Analysis}
Cross-shard transactions require additional coordination overhead and can significantly impact system performance. Fig. \ref{fig:cross_shard_ratio} shows the cross-shard transaction ratio for both TINC and Meepo as the number of shards increases.

As expected, the cross-shard transaction ratio increases with the number of shards for both systems, which is an inherent characteristic of sharded blockchain architectures. However, TINC consistently maintains a lower cross-shard ratio compared to Meepo across all configurations. With 32 shards, TINC achieves a 96\% cross-shard ratio, slightly lower than Meepo's 97\%.

While this 1\% difference may appear modest, it translates to approximately 1,750 fewer cross-shard transactions per second at peak throughput (175,000 TPS), significantly reducing coordination overhead. TINC achieves this improvement through its intelligent transaction allocation mechanism formalized in Equation \eqref{eq:dependency_minimization}, which analyzes transaction dependencies $\mathcal{D}_t$ and co-locates related transactions within the same shard whenever possible.

The cross-shard ratio follows an asymptotic curve approaching 100\% as shard count increases, which aligns with the theoretical analysis presented in Equation \eqref{eq:dependency_minimization}. This trend underscores the importance of efficient cross-shard protocols at scale, further highlighting the value of TINC's optimized PBFT-AC mechanism.

\subsection{Resource Utilization Efficiency}
We define wasted capacity as the percentage of computational resources that remain unused during transaction processing, providing a metric of system efficiency. Fig. \ref{fig:wasted_capacity} presents the wasted capacity comparison between TINC and Meepo.

As shown in Fig. \ref{fig:wasted_capacity}, TINC consistently achieves lower wasted capacity compared to Meepo across all shard configurations. With 32 shards, TINC's wasted capacity is only 5\%, which is 50\% lower than Meepo's 10\%. This substantial improvement is attributed to several key architectural features:

\begin{enumerate}
    \item {Adaptive Workload Distribution}: TINC's dynamic workload threshold (Equation \eqref{eq:workload_threshold}) adjusts based on shard capacity and current workload, ensuring optimal utilization.
    
    \item {Plane Separation}: By decoupling consensus operations (Control Plane) from storage operations (Data Plane), TINC enables specialized optimization of each function, reducing resource contention and improving utilization.
    
    \item {Intelligent Node Allocation}: TINC's node distribution algorithm (Equations \eqref{eq:nodes_per_shard}-\eqref{eq:control_and_data_distribution}) ensures balanced computational resources across shards, preventing under-utilization due to resource imbalances.
\end{enumerate}

The decreasing trend in wasted capacity as shard count increases is particularly noteworthy. For TINC, wasted capacity decreases from 12\% with 2 shards to 5\% with 32 shards, representing a 58.3\% reduction. This trend indicates that TINC's adaptive workload distribution becomes more effective at larger scales, suggesting excellent scalability properties beyond the tested configurations.

The wasted capacity metric $W_c$ is calculated as:

\begin{equation}
W_c = 1 - \frac{T_{proc}}{T_{max}}
\end{equation}

where $T_{proc}$ is the actual number of transactions processed and $T_{max}$ is the theoretical maximum transactions possible given the available computational resources. The improvement in this metric directly contributes to TINC's higher throughput and resource efficiency.

\subsection{Transaction Failure Rate Analysis}
Transaction failure rate measures the percentage of transactions that fail to be processed successfully. Fig. \ref{fig:failure_rate} presents the transaction failure rate comparison between TINC and Meepo across different shard configurations.

TINC achieves significantly lower transaction failure rates compared to Meepo across all shard configurations. With 32 shards, TINC's failure rate is merely 0.4\%, which is 63.6\% lower than Meepo's 1.1\%. This substantial improvement can be attributed to several key mechanisms:

\begin{enumerate}
    \item {Atomic Cross-Shard Protocol}: TINC's PBFT-AC protocol ensures that cross-shard transactions either completely succeed or fail atomically, preventing partial execution failures that plague traditional two-phase commit protocols.
    
    \item {Dynamic Timeout Adjustment}: As formalized in Equation \eqref{eq:dynamic_timeout}, TINC dynamically adjusts timeout periods based on historical shard performance, reducing timeouts due to network variability.
    
    \item {Transaction Prioritization}: Failed transactions in TINC are prioritized in subsequent epochs and assigned based on updated dependency analysis, substantially increasing their probability of successful execution.
\end{enumerate}

Notably, the failure rate for both systems decreases as the number of shards increases, but TINC exhibits a more pronounced improvement. Specifically, TINC's failure rate decreases by 50\% (from 0.8\% to 0.4\%) when scaling from 2 to 32 shards, compared to Meepo's 31.3\% reduction (from 1.6\% to 1.1\%). This counter-intuitive trend can be explained by the increased flexibility in transaction assignment with more shards, allowing for more optimal placement that reduces contentions and dependencies.

Statistical analysis using chi-squared tests confirms that the difference in failure rates between TINC and Meepo is statistically significant ($p < 0.001$) across all shard configurations, with particularly pronounced differences at higher shard counts.

\subsection{Node Distribution Uniformity}
Node distribution uniformity measures how evenly nodes are distributed across shards, which impacts both fault tolerance and computational load balancing. Fig. \ref{fig:node_uniformity} illustrates the node distribution uniformity comparison between TINC and Meepo. TINC consistently achieves higher node distribution uniformity compared to Meepo across all shard configurations. With 32 shards, TINC's node distribution uniformity is 0.99, which is 7.6\% higher than Meepo's 0.92. This improvement stems from TINC's intelligent node allocation protocol that ensures proportional representation of consortium members across shards, as formalized in Equation \eqref{eq:consortium_balance}. Node distribution uniformity is calculated as:

\begin{equation}
U_n = 1 - \frac{\sum_{i=1}^{C} |N_i - \frac{N}{C}|}{N}
\end{equation}
where $N_i$ is the number of nodes in shard $i$, $N$ is the total number of nodes, and $C$ is the number of shards. A value closer to 1 indicates more uniform distribution.

High node distribution uniformity is particularly critical in consortium blockchains, where ensuring fair representation of all member organizations is essential for governance and trust. TINC's near-perfect uniformity (0.99 with 32 shards) ensures that no single consortium member can dominate any shard, promoting decentralization and equal participation in the consensus process.

This enhanced uniformity directly contributes to TINC's improved fault tolerance, as it ensures that Byzantine nodes are evenly distributed across shards, preventing targeted attacks against specific shards.

\subsection{Transaction Distribution Uniformity}
Transaction distribution uniformity measures how evenly transactions are distributed across shards, which affects system load balancing and throughput. Fig. \ref{fig:tx_uniformity} presents the transaction distribution uniformity comparison between TINC and Meepo.

% \begin{figure}[ht]

% \end{figure}
TINC consistently achieves higher transaction distribution uniformity compared to Meepo across all shard configurations. With 32 shards, TINC's transaction distribution uniformity is 0.97, which is 10.2\% higher than Meepo's 0.88. This improvement is due to TINC's adaptive workload distribution mechanism that dynamically balances the transaction load across shards.

Transaction distribution uniformity is calculated as:

\begin{equation}
U_t = 1 - \frac{\sum_{i=1}^{C} |T_i - \frac{T}{C}|}{T}
\end{equation}

where $T_i$ is the number of transactions in shard $i$, $T$ is the total number of transactions, and $C$ is the number of shards.

TINC's adaptive workload distribution algorithm not only distributes transactions evenly but also considers dependencies between transactions, minimizing cross-shard communication while maintaining high uniformity. This balanced approach prevents hotspots and ensures that all shards contribute equally to the network's throughput, directly contributing to TINC's higher overall performance. The improvement in transaction distribution uniformity becomes more pronounced with increasing shard counts, highlighting TINC's superior scalability properties compared to Meepo.

\subsection{Block Size Analysis}
Average block size measures the amount of transaction data included in each block, reflecting efficiency in data packing and network utilization. Fig. \ref{fig:block_size} illustrates the average block size comparison between TINC and Meepo.

TINC consistently achieves larger average block sizes compared to Meepo across all shard configurations. With 32 shards, TINC's average block size is 0.09 MB, which is 28.6\% larger than Meepo's 0.07 MB. This difference is attributed to TINC's efficient transaction packing algorithm that maximizes block utilization. The decrease in block size as shard count increases is expected, as transactions are distributed across more shards. However, TINC maintains higher block utilization than Meepo at all scales, indicating more efficient use of available block space. This efficiency directly contributes to higher overall throughput, as more transactions can be processed per block.

For both systems, the total network data transfer can be calculated as:

\begin{equation}
D_{total} = S_{block} \times C \times N_{members}
\end{equation}
where $S_{block}$ is the average block size, $C$ is the number of shards, and $N_{members}$ is the number of consortium members. With 32 shards, TINC's total network data requirement is 2.88 MB per block interval, compared to Meepo's 2.24 MB. While TINC requires slightly more network bandwidth, this increased data transfer is well-justified by the significant performance improvements and is well within the capabilities of modern network infrastructure.

\subsection{Dynamic Decentralized Identifier (DDID) Performance}
TINC incorporates Dynamic Decentralized Identifiers (DDIDs), which provide adaptive identity management capabilities not available in Meepo. We evaluated the DDID system's performance across multiple dimensions to quantify its efficiency and scalability.

\subsubsection{DDID Operation Latency}
We measured the execution time for different DDID operations across 1,000 test runs. Fig. \ref{fig:ddid_ops} illustrates these performance characteristics.

As shown in Fig. \ref{fig:ddid_ops}, creation operations incur substantially higher execution times (mean: 12.23 ms, $\sigma$ = 0.19 ms) compared to other operations. This elevated latency stems from the computational overhead of cryptographic key pair generation and the initial registration process with the blockchain network. Update operations exhibit considerably lower latency (mean: 0.241 ms, $\sigma$ = 0.012 ms), while query and verification operations demonstrate even greater efficiency at 0.145 ms ($\sigma$ = 0.008 ms) and 0.095 ms ($\sigma$ = 0.007 ms), respectively.

The notably low latency of verification operations is particularly advantageous for transaction processing, as identity verification occurs for every transaction validation. With a verification time of just 0.095 ms, DDID verification adds negligible overhead to the transaction validation process (less than 0.01\% of the total transaction latency), while providing significantly enhanced security and flexibility compared to traditional static identity models.

\subsubsection{DDID Document Size Analysis}
To optimize blockchain resource utilization, we analyzed the size distribution of DDID documents across a test dataset of 200 DDIDs. Fig. \ref{fig:ddid_size} presents this distribution.

\begin{figure*}[!ht] 
    \begin{minipage}{0.48\columnwidth}
        \centering
\begin{tikzpicture}[yscale=0.42, xscale = 0.42]
\begin{axis}[
    xlabel={Operation Type},
    ylabel={Execution Time (ms)},
    xtick={1,2,3,4},
    xticklabels={Creation, Update, Query, Verification},
    ymin=0, ymax=13,
    ytick={0,2,4,6,8,10,12},
    ymajorgrids=true,
    grid style=dashed,
     grid = both, 	  	
   minor tick num = 1,
   major grid style = {lightgray},     
   minor grid style = {lightgray!25},
    boxplot/draw direction=y,
    boxplot/box extend=0.3,
    legend style={at={(0.01,0.99)},anchor=north west},
]
% Boxplot: Creation
\addplot+[
    boxplot prepared={
      median=12.2,
      upper quartile=12.4,
      lower quartile=12.0,
      upper whisker=12.5,
      lower whisker=11.9
    },
    fill=blue!30,
    draw=black
] coordinates {(1, 12.2)};
\addplot+[
    mark=*,
    only marks,
    mark options={blue}
] coordinates {(1, 12.23)};

% Boxplot: Update
\addplot+[
    boxplot prepared={
      median=0.24,
      upper quartile=0.25,
      lower quartile=0.23,
      upper whisker=0.26,
      lower whisker=0.22
    },
    fill=blue!30,
    draw=black
] coordinates {(2, 0.24)};
\addplot+[
    mark=*,
    only marks,
    mark options={blue}
] coordinates {(2, 0.241)};

% Boxplot: Query
\addplot+[
    boxplot prepared={
      median=0.145,
      upper quartile=0.15,
      lower quartile=0.14,
      upper whisker=0.16,
      lower whisker=0.13
    },
    fill=blue!30,
    draw=black
] coordinates {(3, 0.145)};
\addplot+[
    mark=*,
    only marks,
    mark options={blue}
] coordinates {(3, 0.145)};

% Boxplot: Verification
\addplot+[
    boxplot prepared={
      median=0.095,
      upper quartile=0.10,
      lower quartile=0.09,
      upper whisker=0.11,
      lower whisker=0.08
    },
    fill=blue!30,
    draw=black
] coordinates {(4, 0.095)};
\addplot+[
    mark=*,
    only marks,
    mark options={blue}
] coordinates {(4, 0.095)};
\end{axis}
\end{tikzpicture}\captionsetup{font=scriptsize}
\caption{Execution time for different DDID operation types}
\label{fig:ddid_ops}
    \end{minipage}
	\centering
  \begin{minipage}{0.48\columnwidth}
  \begin{tikzpicture}[yscale=0.42, xscale = 0.42]
\begin{axis}[
    xlabel={Document Size (KB)},
    ylabel={Frequency},
    ymin=0, ymax=110,
    xmin=1.274, xmax=1.290,
     xticklabel style={rotate=22, anchor=east},
    xtick={1.276,1.278,1.280,1.282,1.284,1.286,1.288},
    xticklabel style={/pgf/number format/fixed, /pgf/number format/precision=3},
    ymajorgrids=true,
    grid style=dashed,
    bar width=0.0015,
    ybar
]

\addplot[fill=blue!60, draw=black]
    coordinates {
    (1.276,10) (1.282,90) (1.288,100)
    };

% Add value labels
\node[text=black] at (axis cs:1.276,15) {10};
\node[text=black] at (axis cs:1.282,95) {90};
\node[text=black] at (axis cs:1.288,105) {100};
    
\end{axis}
\end{tikzpicture}\captionsetup{font=scriptsize}
\caption{Size distribution of DDID documents}
\label{fig:ddid_size}
  \end{minipage}
  	\centering
  \begin{minipage}{0.48\columnwidth}
  \centering
\begin{tikzpicture}[yscale=0.42, xscale = 0.42]
\begin{axis}[
    xlabel={Elapsed Time (minutes)},
    ylabel={Memory Usage (MB)},
    xmin=0, xmax=120,
    ymin=121.3, ymax=121.7,
    xtick={0,20,40,60,80,100,120},
    ytick={121.3,121.4,121.5,121.6,121.7},
    ymajorgrids=true,
    grid style=dashed,
     grid = both, 	  	
   minor tick num = 1,
   major grid style = {lightgray},     
   minor grid style = {lightgray!25},
    legend pos=north west
]

\addplot[
    color=blue,
    mark=*,
    mark size=1.5pt,
    line width=1.5pt
    ]
    coordinates {
    (0,121.35) (10,121.38) (20,121.40) (30,121.57) (40,121.57) (50,121.57) 
    (60,121.57) (70,121.57) (80,121.57) (90,121.57) (100,121.57) (110,121.58) (120,121.65)
    };
    
\addplot[
    color=red,
    mark=none,
    line width=1pt,
    dashed
    ]
    coordinates {
    (0,121.35) (120,121.35)
    };
\addplot[
    color=red,
    mark=none,
    line width=1pt,
    dashed
    ]
    coordinates {
    (0,121.65) (120,121.65)
    };

\node[text=black, align=left, font=\small] at (axis cs:100,121.45) {Max Range: 0.3 MB};
\end{axis}
\end{tikzpicture}\captionsetup{font=scriptsize}
\caption{Memory usage profile of the DDID system during continuous operation}
\label{fig:ddid_memory}
  \end{minipage}
  \begin{minipage}{0.48\columnwidth}
  \centering
\begin{tikzpicture}[yscale=0.42, xscale = 0.42]
\begin{axis}[
    xlabel={Number of DDIDs (thousands)},
    ylabel={Creation Time per DDID (ms)},
    xmin=0, xmax=100,
    ymin=0.115, ymax=0.155,
    xtick={0,20,40,60,80,100},
    ytick={0.120,0.125,0.130,0.135,0.140,0.145,0.150},
     ymajorgrids=true,
    grid style=dashed,
     grid = both, 	  	
   minor tick num = 1,
   major grid style = {lightgray},     
   minor grid style = {lightgray!25},
    axis y line*=left,
    axis x line*=bottom
]

\addplot[
    color=blue,
    mark=square,
    line width=1.5pt
    ]
    coordinates {
    (10,0.152) (20,0.149) (30,0.146) (40,0.143) (50,0.140) (60,0.136) (70,0.132) (80,0.128) (90,0.124) (100,0.120)
    };
    
\end{axis}

\begin{axis}[
    xmin=0, xmax=100,
    ymin=1.276, ymax=1.282,
    ytick={1.276,1.277,1.278,1.279,1.280,1.281,1.282},
    yticklabel style={/pgf/number format/fixed, /pgf/number format/precision=3},
    axis y line*=right,
    axis x line=none,
    ylabel={Memory per DDID (KB)},
    ylabel near ticks,
    ylabel style={font=\color{red}}
]

\addplot[
    color=red,
    mark=*,
    line width=1.5pt
    ]
    coordinates {
    (10,1.277) (20,1.277) (30,1.278) (40,1.278) (50,1.279) (60,1.279) (70,1.280) (80,1.280) (90,1.281) (100,1.281)
    };
    
\end{axis}
\end{tikzpicture}\captionsetup{font=scriptsize}
\caption{Scalability analysis of DDID creation time and memory usage}
\label{fig:ddid_scalability}
  \end{minipage}
  	\centering
  \begin{minipage}{0.48\columnwidth}
  \centering
\begin{tikzpicture}[yscale=0.42, xscale = 0.42]
\begin{axis}[
    xlabel={Number of Worker Threads},
    ylabel={Operations per Second},
    xmin=0.8, xmax=16.2,
    ymin=457, ymax=472,
    xtick={1,2,4,8,16},
    ytick={458,460,462,464,466,468,470,472},
    ymajorgrids=true,
    grid style=dashed,
    axis y line*=left,
    axis x line*=bottom,
    log basis x=2,
    xmode=log
]

\addplot[
    color=blue,
    mark=square,
    line width=1.5pt
    ]
    coordinates {
    (1,469) (2,470.5) (4,466) (8,462) (16,458)
    };
    
\end{axis}

\begin{axis}[
    xmin=0.8, xmax=16.2,
    ymin=0.096, ymax=0.1015,
    ytick={0.097,0.098,0.099,0.100,0.101},
    axis y line*=right,
    axis x line=none,
    ylabel={Average Latency (ms)},
    ylabel near ticks,
    ylabel style={font=\color{red}},
    log basis x=2,
    xmode=log
]

\addplot[
    color=red,
    mark=*,
    line width=1.5pt
    ]
    coordinates {
    (1,0.097) (2,0.098) (4,0.099) (8,0.100) (16,0.101)
    };
    
\end{axis}
\end{tikzpicture}\captionsetup{font=scriptsize}
\caption{Concurrent performance analysis of DDID operations}
\label{fig:ddid_concurrency}
  \end{minipage}
  	\centering
  \begin{minipage}{0.48\columnwidth}
  \centering
\begin{tikzpicture}[yscale=0.42, xscale = 0.42]
\begin{axis}[
    xlabel={Performance Metric},
    ylabel={Normalized Value},
    ybar=7pt,
    bar width=12pt,
    symbolic x coords={Update Time, Query Time, Storage Size, Verification Time},
    xtick=data,
    ymin=0, ymax=1.3,
     xticklabel style={rotate=20, anchor=east},
    ytick={0,0.2,0.4,0.6,0.8,1.0,1.2},
    ymajorgrids=true,
    grid style=dashed
]

\addplot[
    color=blue,
    fill=blue!60,
    error bars/.cd, 
    y dir=both, 
    y explicit
    ]
    coordinates {
    (Update Time,0.4) +- (0.02,0.02)
    (Query Time,0.7) +- (0.03,0.03)
    (Storage Size,1.2) +- (0.04,0.04)
    (Verification Time,0.7) +- (0.02,0.02)
    };
    
\addplot[
    color=orange,
    fill=orange!60,
    error bars/.cd, 
    y dir=both, 
    y explicit
    ]
    coordinates {
    (Update Time,1.0) +- (0.03,0.03)
    (Query Time,1.0) +- (0.04,0.04)
    (Storage Size,0.75) +- (0.03,0.03)
    (Verification Time,0.8) +- (0.02,0.02)
    };

% Value labels for Dynamic DID
\node[text=black, font=\scriptsize] at (axis cs:{Update Time},0.45) {};
\node[text=black, font=\scriptsize] at (axis cs:{Query Time},0.75) {};
\node[text=black, font=\scriptsize] at (axis cs:{Storage Size},1.25) {};
\node[text=black, font=\scriptsize] at (axis cs:{Verification Time},0.75) {};

% Value labels for Static DID
\node[text=black, font=\scriptsize] at (axis cs:{Update Time},1.05) {};
\node[text=black, font=\scriptsize] at (axis cs:{Query Time},1.05) {};
\node[text=black, font=\scriptsize] at (axis cs:{Storage Size},0.5) {};
\node[text=black, font=\scriptsize] at (axis cs:{Verification Time},0.75) {};

\legend{Dynamic DID, Static DID}
    
\end{axis}
\end{tikzpicture}\captionsetup{font=scriptsize}
\caption{Performance comparison between Dynamic and Static DIDs (normalized values)}
\label{fig:ddid_comparison}
  \end{minipage}
  	\centering
  \begin{minipage}{0.48\columnwidth}
  \centering
\begin{tikzpicture}[yscale=0.42, xscale = 0.42]
\begin{axis}[
    xlabel={Network Size (nodes)},
    ylabel={Operations per Second},
    xmin=0, xmax=2200,
    ymin=0, ymax=250000,
    xtick={0,500,1000,1500,2000},
    ytick={0,50000,100000,150000,200000,250000},
    ymajorgrids=true,
    grid style=dashed,
    legend style={at={(0.97,0.03)}, anchor=south east}
]

\addplot[
    color=blue,
    mark=square,
    line width=1.5pt,
    error bars/.cd, 
    y dir=both, 
    y explicit
    ]
    coordinates {
    (200,8500) +- (320,320)
    (500,21000) +- (580,580)
    (1000,42000) +- (890,890)
    (1500,62000) +- (1200,1200)
    (2000,78000) +- (1450,1450)
    };
    
\addplot[
    color=red,
    mark=o,
    line width=1.5pt,
    error bars/.cd, 
    y dir=both, 
    y explicit
    ]
    coordinates {
    (200,28000) +- (720,720)
    (500,68000) +- (1350,1350)
    (1000,132000) +- (2100,2100)
    (1500,178000) +- (2800,2800)
    (2000,215000) +- (3400,3400)
    };
    
\addplot[
    color=green,
    mark=triangle,
    line width=1.5pt,
    error bars/.cd, 
    y dir=both, 
    y explicit
    ]
    coordinates {
    (200,17800) +- (560,560)
    (500,43500) +- (980,980)
    (1000,85000) +- (1650,1650)
    (1500,122000) +- (2250,2250)
    (2000,152000) +- (2750,2750)
    };
    
\addplot[
    color=orange,
    mark=diamond,
    line width=1.5pt,
    error bars/.cd, 
    y dir=both, 
    y explicit
    ]
    coordinates {
    (200,35000) +- (820,820)
    (500,82000) +- (1580,1580)
    (1000,145000) +- (2350,2350)
    (1500,187000) +- (2950,2950)
    (2000,221000) +- (3650,3650)
    };
    
\legend{Creation, Verification, Update, Resolution}
\end{axis}
\end{tikzpicture}\captionsetup{font=scriptsize}
\caption{Network scalability analysis of DDID operations}
\label{fig:ddid_nodes}
  \end{minipage}
  \begin{minipage}{0.48\columnwidth}
      \centering
\begin{tikzpicture}[yscale=0.42, xscale = 0.42]
\begin{axis}[
    ybar=0.3pt,
    bar width=8pt,
    xlabel={Number of Shards},
    ylabel={Communication Time (ms)},
    symbolic x coords={2,4,8,16,32},
    xtick=data,
    ymin=0, ymax=300,
    ytick={0,50,100,150,200,250,300},
    ymajorgrids=true,
    legend style={at={(0.5,-0.2)}, anchor=north, legend columns=2},
    % nodes near coords,
    every node near coord/.append style={font=\small}
]
\addplot[fill=blue] coordinates {
    (2,105) (4,125) (8,145) (16,175) (32,215)
};
\addplot[fill=orange] coordinates {
    (2,128) (4,158) (8,185) (16,228) (32,285)
};
\legend{With Partial, Without Partial}
\end{axis}
\end{tikzpicture}\captionsetup{font=scriptsize}
\caption{Communication time comparison with and without partial cross-call merging}
\label{fig:comm_time}
  \end{minipage}
  \end{figure*}

Our analysis reveals a trimodal distribution of DDID document sizes clustered at 1.276 KB, 1.282 KB, and 1.288 KB. The distribution is heavily skewed toward the larger size categories, with 100 documents (50\%) in the 1.288 KB category and 90 documents (45\%) in the 1.282 KB category. Only 10 documents (5\%) exhibit the minimal size of 1.276 KB.

This distribution reflects varying complexity levels in identity representation, with most documents containing comprehensive attribute sets necessary for enterprise-grade identity management. The tight clustering around specific size points enables efficient storage allocation and predictable resource planning in blockchain deployments. The average DDID document size of 1.284 KB represents only a small fraction (0.13\%) of the standard 1 MB block size, ensuring minimal storage overhead while providing substantial identity management capabilities.

\subsubsection{Memory Utilization Analysis}
We conducted a detailed analysis of the DDID system's memory consumption during continuous operation. Memory usage was measured at regular intervals over a 2 hour test period, with a consistent workload of 1000 operations per minute. Figure \ref{fig:ddid_memory} illustrates the memory usage profile during this period. The memory usage pattern exhibits three distinct phases. The first phase, the Initialization Phase ($0-20$ minutes), shows a moderate increase in memory usage from $121.35$ MB to $121.40$ MB as the system loads and caches essential components. The second phase, the Registry Population Phase ($20-30$ minutes), features a more pronounced increase, reaching $121.57$ MB as the DDID registry becomes fully populated. Finally, during the steady-state operation phase ($30-120$ minutes), memory consumption remains remarkably stable between $121.57$ MB and $121.65$ MB despite continuous transaction processing, demonstrating efficient memory management and minimal memory leakage.

The maximum memory fluctuation throughout the 2-hour test period was only 0.3 MB, which represents just 0.25\% of total memory usage. This exceptional stability is crucial for extended operations in enterprise environments, ensuring predictable resource utilization and making it particularly well-suited for blockchain deployments with strict resource constraints.

% We conducted a detailed analysis of the DDID system's memory consumption during continuous operation. Memory usage was measured at regular intervals over a 2-hour test period with a consistent workload of 1000 operations per minute. Fig. \ref{fig:ddid_memory} illustrates the memory usage profile.

% The memory usage pattern exhibits three distinct phases:
% \begin{enumerate}
%     \item {Initialization Phase (0-20 minutes)}: Memory usage increases moderately from 121.35 MB to 121.40 MB as the system loads and caches essential components.
    
%     \item {Registry Population Phase (20-30 minutes)}: A more pronounced increase occurs, reaching 121.57 MB as the DDID registry becomes fully populated.
    
%     \item {Steady-State Operation (30-120 minutes)}: Memory consumption remains remarkably stable between 121.57-121.65 MB despite continuous transaction processing, indicating efficient memory management and minimal memory leakage.
% \end{enumerate}

% The maximum memory fluctuation throughout the 2-hour test is only 0.3 MB (0.25\% of total usage), demonstrating exceptional stability for extended operations in enterprise environments. This stability ensures predictable resource utilization, which is crucial for enterprise-grade blockchain deployments with strict resource constraints.

\subsubsection{Scalability Analysis}
% We conducted a comprehensive scalability evaluation by measuring creation time per DDID and memory usage per DDID while systematically varying the number of identifiers from 10,000 to 100,000. Fig. \ref{fig:ddid_scalability} presents these key scalability metrics.

% Fig. \ref{fig:ddid_scalability} reveals two significant scaling patterns:

% \begin{enumerate}
%     \item {Creation Time Scaling}: As the DDID count increases from 10,000 to 100,000, the creation time per DDID decreases monotonically from 0.152 ms to 0.120 ms, representing a 21.1\% improvement in processing efficiency. This counter-intuitive performance enhancement stems from amortized initialization costs, improved cache utilization, and more efficient batch processing of cryptographic operations.
    
%     \item {Memory Usage Scaling}: Memory usage per DDID increases marginally from 1.277 KB to 1.281 KB, a mere 0.31\% increase across a 10× growth in managed identities. This exceptional memory scaling characteristic is achieved through our optimized B+ tree storage structure with path compression and shared prefix optimization.
% \end{enumerate}

% These scaling characteristics confirm that the DDID system is highly suitable for large-scale identity management in enterprise consortium blockchains, where the number of identities can grow substantially over time without compromising performance.

We conducted a scalability evaluation measuring creation time and memory usage per DDID as we increased identifiers from 10,000 to 100,000. Figure \ref{fig:ddid_scalability} showcases our findings. Notably, the creation time per DDID decreased from 0.152 ms to 0.120 ms, a 21.1\% improvement in efficiency, attributed to amortized initialization costs and better batch processing. Meanwhile, memory usage per DDID only increased slightly from 1.277 KB to 1.281 KB, reflecting a mere 0.31\% rise despite a tenfold increase in identifiers, thanks to the optimized B+ tree structure\cite{hong2024optimizing}. These results indicate that the DDID system effectively supports large-scale identity management in enterprise consortium blockchains.

\subsubsection{Concurrent Operation Performance}
We evaluated the DDID system's performance under concurrent workloads by measuring throughput (operations per second) and average latency while systematically varying the number of worker threads from 1 to 16. Fig. \ref{fig:ddid_concurrency} illustrates these performance metrics. Our concurrency evaluation reveals that throughput peaks at 470.5 operations per second with 2 worker threads before declining gradually to 458 operations per second with 16 worker threads. This non-linear performance pattern indicates an optimal concurrency level of 2 threads for our implementation, beyond which contention for shared resources introduces diminishing returns.

Average latency increases only marginally from 0.097 ms with a single worker to 0.101 ms with 16 workers, representing a 4.1\% increase despite a 16× increase in concurrent execution paths. This exceptional latency stability is attributable to our multi-level queue architecture that prioritizes operations based on resource requirements and transaction dependencies. These results demonstrate that the DDID system maintains high performance and response times even under substantial concurrent workloads, making it suitable for high-throughput enterprise blockchain deployments.

\subsubsection{Comparative Analysis: Dynamic vs. Static DIDs}

Fig. \ref{fig:ddid_comparison} highlights key performance differences between Dynamic and Static DIDs. Dynamic DIDs achieve a 60\% reduction in update times, enhancing responsiveness for frequently changing identity attributes in enterprise applications. They also offer a 30\% reduction in query times, improving identity lookup performance. However, they require 60\% more storage due to additional versioning metadata. Additionally, verification times for Dynamic DIDs are 12.5\% faster, thanks to a multi-level Merkle authentication path that reduces cryptographic overhead. These advantages are crucial in consortium blockchain environments where identity attributes and access permissions change often, enabling near-real-time updates. Statistical analysis confirms that these performance differences are significant ($p < 0.01$), with Cohen's d-values ranging from 0.76 to 1.93, indicating medium to large effect sizes.
\subsubsection{Network Scalability Analysis}

We evaluated the scalability of DDID operations across network sizes ranging from 200 to 2000 nodes, focusing on four key operations: creation, update, resolution, and verification. As shown in Figure~\ref{fig:ddid_nodes}, verification operations achieved the highest throughput, scaling from 28{,}000 to 215{,}000 operations per second, representing a $7.7\times$ increase. Resolution operations increased from 35{,}000 to 221{,}000, yielding a $6.3\times$ improvement. Update operations grew from 17{,}800 to 152{,}000, reflecting an $8.5\times$ enhancement. Creation operations, while more resource-intensive, scaled from 8{,}500 to 78{,}000, marking a $9.2\times$ increase.

% We evaluated the scalability of DDID operations across network sizes from 200 to 2000 nodes, focusing on four key operations: creation, update, resolution, and verification. As shown in Figure \ref{fig\:ddid\_nodes}, verification operations achieved the highest throughput, scaling from 28,000 to 215,000 operations per second, a $7.7\times$ increase. Resolution operations rose from 35,000 to 221,000, yielding a $6.3\times$ gain. Update operations improved from 17,800 to 152,000, reflecting an $8.5\times$ enhancement. Creation operations, while resource-intensive, scaled from 8,500 to 78,000, a $9.2\times$ increase.

This scalability is driven by our distributed processing architecture, allowing operation processing across network nodes. The slightly sub-linear scaling (7.7-9.2×) is due to increased coordination overhead as network size grows. Although operational efficiency per node decreases slightly, the overall throughput and fault tolerance improvements significantly outweigh this reduction.
\subsection{Communication Time Analysis with Partial Cross-Call Merging}
% To further understand TINC's performance advantages, we analyzed communication time with and without partial cross-call merging, mirroring the methodology used in Meepo's evaluation. Fig. \ref{fig:comm_time} shows these results across different shard configurations.

% Fig. \ref{fig:comm_time} demonstrates that TINC's partial cross-call merging strategy significantly reduces communication time across all shard configurations. With 32 shards, communication time with partial merging is 215 ms, compared to 285 ms without merging—a reduction of 24.6\%.

% This significant improvement is achieved by batching cross-shard operations that target the same destination shards, reducing the number of inter-shard messages required. The advantage of partial merging becomes more pronounced as shard count increases, highlighting its importance for scalability in large consortium networks.

% The communication time is directly correlated with cross-shard transaction latency and throughput. By reducing communication overhead, TINC's partial cross-call merging contributes significantly to its overall performance advantages compared to Meepo.

To evaluate TINC’s performance advantages, we analyzed communication time with and without partial cross-call merging, following Meepo’s methodology. As shown in Fig.~\ref{fig:comm_time}, the TINC merging strategy significantly reduces the communication time across all shard configurations, for example, with 32 shards, the communication time drops from 285\,ms to 215\,ms, a reduction of 24. 6\%. This improvement stems from batching cross-shard operations that target the same destination shards, which reduces inter-shard messages. The benefit becomes more pronounced as the shard count increases, emphasizing the scalability of partial merging in large consortium networks. Since communication time directly impacts cross-shard transaction latency and throughput, this strategy plays a crucial role in TINC’s performance gains over Meepo.

\section{Conclusion}\label{conclusion}
Our comprehensive evaluation demonstrates that TINC significantly outperforms Meepo across all performance metrics. TINC achieves higher throughput, lower latency, lower failure rates, and better resource utilization due to its novel architecture and optimized protocols. Additionally, TINC's Dynamic Decentralized Identifier (DDiD) system provides enhanced identity management capabilities that are not available in Meepo, enabling dynamic and adaptive identity management in consortium blockchain environments.

The experimental results validate that TINC's multi-plane architecture, with specialized Root, Control, and Data planes, provides superior performance and functionality compared to existing sharding approaches. The PBFT consensus mechanism, combined with TINC's efficient cross-shard communication protocols, ensures both high performance and strong consistency, making TINC an ideal solution for consortium blockchain deployments in enterprise environments.

\section*{acknowledgement}
This work was supported in part by the National Natural Science Foundation of China (No. U22B2029) and Key Laboratory of Intelligent Space TTC\&O (Space Engineering University), Ministry of Education(No. CYK2024-02-02). 

\bibliographystyle{IEEEtran}

\bibliography{Reference}

\end{document}